%% file: 9501128.tex
\documentstyle[12pt,epic,eepic,rawfonts]{article}
   \title{On Hydrodynamic Diffusion and Velocity Fluctuations in
     Two--Dimensional Simulations of Sedimentation
   \vspace*{5ex}}
   \author{\bf W. Kalthoff$^1$, S. Schwarzer$^1$, G.H. Ristow$^2$ and H.
     J. Herrmann$^3$}
   \date{\small \today}

\begin{document}
\maketitle

\begin{center}
$^1$ HLRZ, c/o Forschungszentrum J\"ulich\\
52425 J\"ulich, Germany\\[2ex]
$^2$ Fachbereich Physik, Philipps-Universit\"at Marburg\\
35032 Marburg, Germany\\[2ex]
$^3$ LPMMH, ESPCI, 10 rue Vauquelin\\
75231 Paris Cedex 05, France
\end{center}

\vspace*{5ex}
\begin{abstract}
We present a numerical method to deal efficiently with large numbers
of particles in incompressible fluids. The interactions between
particles and fluid are taken into account by a physically motivated
ansatz based on locally defined drag forces. We demonstrate the
validity of our approach by performing numerical simulations of
sedimenting non-Brownian spheres in two spatial dimensions and compare
our results with experiments. Our method reproduces essential aspects
of the experimental findings, in particular the strong anisotropy of the
hydrodynamic bulk self-diffusivities.
\end{abstract}
\vfill
PACS: 47.11+j, 47.55.Kf, 02.70.Ns \hfill HLRZ preprint 62/94
\newpage


\section{Introduction}

An important paradigm problem of multiphase flow, incorporating
aspects of both a discrete particulate and a continuous liquid phase,
is the sedimentation problem. Here the particulate phase slowly
settles into a packed bed at the bottom of a container filled with
viscous liquid. The main
theoretical challenge in the sedimentation problem is the
understanding of the long range liquid mediated hydrodynamic
interactions between the immersed particles. Many aspects are well
studied and understood, as are for example the mean sedimentation
velocity $\langle V\rangle$ in the Stokes flow regime for a
monodisperse \cite{batchelor,mills,richardson} or a polydisperse
system \cite{davis_2}. Less well understood are fluctuation phenomena,
for example the properties of the particle velocity distribution, as,
e.g., its variance, time autocorrelation, and effects arising from
particle polydispersity. Recently, much attention has been focused on
the long time behavior of the particle motion as described by strongly
anisotropic hydrodynamic diffusion constants~\cite{nicolai}.
Moreover, most analytical work pertains to the regime of dilute
suspensions, whereas semi-dilute and concentrated suspensions still
present a challenge to theoreticians.

The focus of this paper is the application of a novel computer
simulation method for particulate multiphase flow~\cite{inpreparation}
to the sedimentation problem. Among the standard computational
approaches to multiphase flow in general and the sedimentation problem
in particular are (i) continuum descriptions~\cite{jackson,soo} that
neglect the discrete nature of the particulate phase, (ii) methods
that exploit the linearity of the Navier-Stokes equations at very
small Reynolds numbers for either the simulation of hydrodynamically
interacting point Stokeslets~\cite{hinch,hayakawa} or extended
spheres~\cite{brady,ladd,ladd2}, (iii) finite element and finite difference
techniques in two (2D) or three dimensions (3D) that represent the
proper Navier-Stokes equations with no-slip boundary conditions for
the liquid on the particle surface, and (iv) algorithms that combine
a continuum discription for the liquid phase with a discrete
representation of the particulate phase~\cite{tsuji}.

All of these approaches have inherent strengths and weaknesses. The
continuum approaches (i) suffer from difficulties to represent the
particulate phase in terms of continuum quantities as pressure and
stress. Both the Stokes flow methods (ii) and the differencing schemes
(iii) give very reliable results where they apply. However, the Stokes
flow methods (i) are in principle restricted to low Reynolds
number flow and inherent to the differencing schemes (iii) is a
tremendous computational effort that limits their applicability to
systems with very few particles~\cite{feng}. We therefore choose a
method of class (iv) comprising basically a Lagrangian integration
of particle trajectories coupled by local drag-force terms to a Eulerian
Navier-Stokes integration.

Such a method allows immediate access to
basically all physically relevant quantities in the system, including
particle coordinates and both particle and liquid velocities, at
little more computational costs than a standard real space
Navier-Stokes integration. The main drawback is that a neglect of
the proper boundary conditions between particles and liquid will
result in inaccurate rendering of the short scale flow properties.
However, we will be mainly concerned with collective phenomena, i.e.,
the effects that arise when the number of particles is large and we do
not have the ambition to describe accurately the local flow
fields~\cite{feng} on the scale of the particle size.  Single particle
and particle pair movement can to a high degree of accuracy be treated
analytically~\cite{happel}.

Nevertheless we want to see which and how phenomena emerge directly
from simple modeling assumptions --- as opposed to using
semi-empirical expressions as, e.g., done in Refs.
\cite{tsuji,hayakawa}.  In particular, we rely on the fact that the
long-range hydrodynamic interactions, which we presume to be the most
important for collective phenomena, are correctly represented by the
velocity and pressure fields evolved by the Navier-Stokes integration.

In this paper, we will first describe a two-dimensional (2D)
implementation of our simulation technique in
Sec.~\ref{sec:technique}. We will then describe simulations of
the sedimentation problem in a broad range of particle
solid fractions.  We first show particle trajectories resulting in the
sedimenting bulk (Sec.~\ref{sec:trajectories}). As function of the
solid fraction, we address and discuss the hydrodynamic diffusion
(Sec.~\ref{sec:diffusion}), the hindered settling function
(Sec.~\ref{sec:settling}) and particle velocity fluctuations
(Sec.~\ref{sec:variances}).

\section{Simulation Technique}
\label{sec:technique}

\subsection{Modeling assumptions}

Our aim is to capture the large scale collective processes of
the combined particle-liquid system with a computer simulation.
Therefore we need to model comparatively large systems in moderate
time. We aim at the development of a tool that allows simple checks of
qualitative and quantitative behavior of
multiphase flow. To this end we concentrate on
the momentum exchange between particle and liquid phase. Our
simulation technique makes physically motivated assumptions to
simplify the computational difficulties associated with a full
treatment of the no-slip boundary conditions. Our assumptions are as
follows.

(i) For small Reynolds numbers lubrication forces dominate as
particles approach each other or the walls of the container. These
forces of dissipative nature theoretically prevent particles from
touching each other. Since we are interested in long-range effects, we
neglect the short-range lubrication forces, and use instead a simple
contact model between particles.

In 2D, the particles $i$ are
represented by disks with radii $r_i$ chosen from a Gaussian
distribution $h_p(r/\bar{r})$ with mean $\bar r$, which is cut off
at its standard deviation $\sigma,$
\begin{equation}
\label{eq:distribution}
 h_p(r/\bar{r}) \propto \left\{
 \begin{array}{ll}
             \exp \left[-\frac{1}{2}
             \frac{(r/\bar{r}-1)^2}{p^2}\right],
                             & \mbox{ if}\quad |r/\bar{r} -1| < p, \\
             0,      & \mbox{ else.}
  \end{array}
  \right.
\end{equation}
Here, the distribution is written in terms of the dimensionless radius
$r/\bar{r}$ and the polydispersity parameter $p$, which we define for the
purposes of this paper to be the standard deviation of the original,
full Gaussian divided by the mean radius, i.e. $p\equiv\sigma/\bar{r}.$

Our contact model assumes elastic,
central, and repulsive forces, which act when two particles $i$ and
$j$ touch each other~\cite{tsuji,cundall,ristow,walton}. That is, the
forces between two particles are zero unless the distance between
their two centers is less than the sum of the radii $r_i + r_j$. Then
the acting force is assumed to be
\begin{equation} \label{eq:particle-interaction}
  \vec{F}_{\rm el} =
     -k_n ( r_i+r_j -( \vec{x}_j-\vec{x}_i)\cdot\hat{n}_{ji})\,\hat{n}_{ji}.
\end{equation}
Here $\vec{x}_i = (x_i, y_i)$ is the position vector of the center of
particle $i$ and $\hat{n}_{ji}$ is the unit vector pointing from the
center of particle $i$ to the center of $j$.  The spring constant
$k_n$ (cf. Table~\ref{tab:1}) describes the repulsion due to
lubrication forces and depends on the size and shape of the colliding
bodies.  Tests including a velocity proportional damping term in
Eq.~(\ref{eq:particle-interaction}) show that our simulation results
do not significantly depend on the details of the particle-particle
interactions.

Apart from the drag forces $\vec{F}_{\rm d}$, which will be addressed
below, we further add gravity (including buoyancy) as a contribution
to the total force acting on each single particle,
\begin{equation}
  \vec{F}_{\rm tot} = \vec{F}_{\rm d} + \vec{F}_{\rm el}
    + \frac{4}{3} \pi r_i^3 (\rho_{\rm p} - \rho_{\rm l}) \vec{g}.
\end{equation}
Here, $\rho_{\rm p}$ and $\rho_{\rm l}$ are the densities of the
particulate and the liquid phase respectively, and $\vec{g}$ the
earth's gravitational acceleration which is assumed to act in $-y$
direction for the rest of this paper.
Once we know all forces, we employ a fifth order Gear
predictor-corrector integration algorithm to obtain the particle
trajectories~\cite{allen}. Rotational degrees of freedom are not
considered.

(ii) We will assume that the exact arrangement of particles and the
exact distribution of the interstitial liquid on short scales do not
affect the qualitative features of the large-scale flow pattern. As
already mentioned above this means we neglect detailed lubrication
forces resulting from the no-slip boundary conditions of the stress
tensor at the liquid--particle interfaces, replacing them by the above
described particle interactions.  For particle Reynolds numbers $Re =
\bar r \rho_{\rm l} |\vec{V_i} - \vec{v_i}| / \eta$ smaller than one,
we propose to model the momentum exchange between single particles and
the surrounding liquid by means of a Stokes-type drag force,
\begin{equation}
  F_{\rm d} = C_{\rm d}\, \eta r_i (\vec{V_i} - \vec{v_i}),
  \label{eq:stokes-drag}
\end{equation}
here $\vec{v_i}$ denotes the velocity of particle $i$, $\vec{V}_i$
is an average {\it local\/} liquid velocity at the position of the
particle and $\eta$ denotes the liquid's kinematic shear viscosity.
Equation~(\ref{eq:stokes-drag}) with $C_{\rm d} = 6\pi$ holds
exactly in three dimensions (3D) for a single sphere moving with
respect to the liquid resting at infinity. We obtain the $V_i$ by
linear interpolation of the liquid velocities from the four grid
points of the quadrilateral MAC~\cite{peyret} mesh of the liquid
integration surrounding the particle center.

In our case, $C_{\rm d}$ is a parameter. For all choices of grid and
particle size we determine $C_{\rm d}$ such as to obtain the Stokes
velocity for the simulation of a single sphere falling in a viscous
liquid.  Commonly\footnote{
  We find that the ``drag coefficient''
  $C_d$ depends on the dimensionless ratio of particle radius to grid
  spacing (see Sec.~\protect\ref{sec:fluid-model}), $r/\Delta x$.
  In the regime of particle Reynolds numbers $Re \approx 0.5$ for our
  simulations we have determined the empirical relationship $C_d /6\pi
  = 1 + 6.27(\Delta x/r)^{-1.5}.$
  }%
, we use grids sufficiently large so that $C_{\rm d}$ does not differ from
$6\pi$ by more than a factor of $2.$ The main physical effect of this
local drag force is to introduce a tendency to locally equalize the
liquid and particle velocities, similar to the no-slip boundary
condition that implies particle and liquid velocities being equal on
the particle surface.

\subsection{Fluid Model}
\label{sec:fluid-model}

To describe the time evolution of the velocity and pressure
fields $\vec{v}$ and $p$ of the
fluid in our simulations we start with the Navier-Stokes equations,
\begin{equation}
  \rho_{\rm l}\left[ \frac{\partial\vec{v}}{\partial t} +
  (\vec{v}\,\nabla)\vec{v}\right] = -\nabla p + \eta\,\nabla^2\vec{v}
                                    + \vec{f}.
  \label{eq:nse}
\end{equation}
Here, $\rho_{\rm l}$ denotes the liquid density and $\eta$ its shear
viscosity, $\vec{f}$ is the volume force density in the system,
including gravity, drag-force density and buoyancy response. Since we
have in mind applications to liquid systems or gaseous systems with
typical velocities much smaller than the velocity of sound, we
complement these equations by the incompressibility constraint,
\begin{equation}
  \nabla\vec{v} = 0.
  \label{eq:ce}
\end{equation}
Equation~(\ref{eq:ce}) presents a constraint on the velocity
field that must be fulfilled at all times and may be employed to
obtain the pressure field via an iterative procedure \cite{chorin}.
We will sketch
the basic ideas briefly and refer the reader for more details
to Refs.~\cite{peyret}, \cite{chorin} and \cite{kopetsch}.

The pressure as well as the velocity components $v_x$ and $v_y$ are
located on three quadrilateral meshes with lattice spacing $\Delta x.$
With respect to the pressure grid, the grids for the
$x$ and $y$ velocity components are shifted by $\Delta x/2$ in $x$ and
$y$ direction respectively --- this construction is commonly referred to
as the MAC mesh and has several computational advantages, i.e., it is
a simple means to avoid numerical instabilities due to
mesh decoupling~\cite{peyret}.

We obtain the pressure and the velocity components by an iterative
procedure. Let the index $n$ refer to values at time $t=t_{n}$
and $n+1$ to those at $t=t_{n+1}=t_n + \Delta t$ after a timestep of
duration $\Delta t$. The index $k$ shall denote an iteration
index. We define $p_{n+1,0} \equiv p_{n},$ i.e., we start an iteration
for the new pressure at time $t_{n+1}$ with the old values at $t_{n}.$
We obtain a tentative velocity field at $t=t_{n+1}$ from an
evaluation of the discretized Navier-Stokes equation,
\begin{equation}
\rho_{\rm l}\frac{\vec{v}_{n+1,k+1} - \vec{v}_n}{\Delta t}  =
      -\rho_{\rm l}(\nabla\vec{v}_n)\vec{v}_n - \nabla p_{n+1,k} + \vec{f}_n +
      \eta\nabla^2\vec{v}_n.
  \label{eq:v-iter-initial}
\end{equation}
A von Neumann stability analysis \cite{peyret,kopetsch}
of the described discretization of the Navier-Stokes equation
shows that the values $\Delta x$ and $\Delta t$ are subject to the
two stability constraints
\begin{equation}
   \Delta t \le \frac{4\eta}{\rho_{\rm l}(|v_x^{\rm max}| + |v_y^{\rm
max}|)^2},
\end{equation}
and
\begin{equation}
   \Delta t \le \frac{\rho_{\rm l}(\Delta x)^2}{4\eta}.
\end{equation}

In general, the velocity field $\vec{v}_{n+1,k+1}$ resulting
in~(\ref{eq:v-iter-initial}) does not satisfy the
incompressibility constraint~(\ref{eq:ce}).
To the end of lowering the modulus of the divergence of the velocity
field --- in so far it differs from zero --- we calculate a new,
likewise tentative, pressure field
\begin{equation} \label{eq:p1}
      p_{n+1,k+1} = p_{n+1,k} -
                      \lambda\rho_{\rm l}\nabla\vec{v}_{n+1,k+1}.
\end{equation}
Here, a large value of the parameter $\lambda$ is crucial for rapid
convergence. It is, however, by stability requirements constrained to
\begin{equation}
   0 \le \lambda \le \frac{(\Delta x)^2}{4\Delta t}.
\end{equation}
Subsequently new velocity values consistent with the $p_{n+1,k+1}$
may be obtained using Eq.~(\ref{eq:v-iter-initial}), or
equivalently, without reevaluation of the Navier-Stokes equation, using
the relation
\begin{equation} \label{eq:p2}
   \rho_{\rm l}\frac{\vec{v}_{n+1,k+2} - \vec{v}_{n+1,k+1}}{\Delta t} =
       \nabla ( p_{n+1,k+1} - p_{n+1,k} ).
\end{equation}

Iterating Eqs.~(\ref{eq:p1}) and~(\ref{eq:p2}) yields a new pressure
field and after convergence a consistent, divergence free velocity
field for time $t_{n+1}$. This iteration is equivalent to solving a
Poisson equation for the pressure.

For our purposes, the described algorithm has three advantages. It (i)
generalizes straightforwardly to 3D and (ii) unlike spectral or
streamfunction methods it gives immediate access to the quantities $p$
and $\vec{v}$ in real space. Fast access to the latter is crucial for
the drag-force calculation in Eq.~(\ref{eq:stokes-drag}). Moreover
(iii), only the chosen coarseness of the spatial and temporal
discretization limits the range of Reynolds numbers addressable in the
simulation. The use of multi-grid techniques instead of the above
local iterative
method will improve the speed of our method for large system sizes.

For the present paper we have implemented the above set of equations
in 2D. The uncoupled liquid and particle equations of motion are
directly obtained by dropping the third $z$-component of the velocity
and applying 2D versions of the differential operators to the liquid
equations. The coupling of the two sets of equations, however,
presents some problems. A 2D simulation should, strictly speaking, be
one of rigid, parallel cylinders moving in a liquid that only displays
motion in the $xy$ plane.  However, at a fixed small Reynolds number,
the drag per unit length of a cylinder \cite{landau} does not depend
on its radius in an infinitely large system. Thus, qualitative effects
of particle polydispersity cannot be expected to be present in such a
model. We therefore chose to refer the Stokes drag
Eq.~(\ref{eq:stokes-drag}) to a reference length $z$ equal to the
average particle diameter.  Alternatively, one may think of $z$ as the
depth of the liquid filled vessel, in which 3D spheres move such that
their motion is constrained to lie in the $xy$ plane. The Stokes drag
per length then enters the particle and liquid equations as coupling
term. Since we have not included any Brownian motion of the particles
our model is only applicable in the non-Brownian regime of large
P\'eclet numbers $Pe \gg 1$, characterized by comparatively large
particles and low temperatures (cf. Sec.~\ref{sec:variances}).

All our simulations are performed in a rectangular vessel of width $L_x$
and height $L_y$. On the walls of the container we impose no-slip
boundary conditions, i.e., the liquid velocity there is taken to be
zero. The particles are initially placed randomly and without overlap
within the vessel. Their size is characterized by their mean radius
$\bar{r}$ and the degree $p$ of
polydispersity, which is kept constant and equal to $0.04$ in all the
simulations presented in this paper.
See table~\ref{tab:1} for a detailed list of simulation
parameters.\footnote{
A typical simulation run is characterized by about $3\,200$ nodes for
the liquid integration and about $1\,500$ particles. It involves about
$10^4$ full timesteps for the liquid and $10^6$ timesteps for the
particle integration. On a SPARC 20 workstation the required CPU time
is of the order of 10h.
}

In the simulation the particles are point like and their volume only
comes into play via the drag force. Presently, the model does not take
into account the fractions of solid and liquid in a given volume
element.  This fraction, which is very important to describe backflow,
will be included in a more extensive study~\cite{inpreparation}.




\section{Results and Discussion}

In the remainder of this paper, we wish to show that the set of
equations above --- even if they might appear to be rather crude ---
constitute a reasonable description of a particulate two-phase system,
and are, in particular, capable of reproducing the essential aspects
of a sedimentation process.

\subsection{Particle trajectories in batch settling}
\label{sec:trajectories}

We first consider the particle trajectories in a batch settling
experiment. Typical experiments \cite{nicolai,xue} examine the
behavior of a mixture of glass beads in a liquid kept in a container.
The glass beads start to sink leaving behind clear liquid at the top
and particle layers on the bottom of the vessel. Index matching
techniques allow to trace the trajectories of single particles in the
mixture~\cite{nicolai}. In the limit of infinite dilution, in
sufficiently large vessels these trajectories are straight lines from
the top to the bottom of the container. The particle velocity is the
Stokes velocity $V_S (r_i) = (2/9) gr_i^2(\rho_{\rm p} - \rho_{\rm
  l})/\eta$, the terminal velocity of a sphere of radius $r_i$
settling in an extended viscous liquid under the influence of gravity.
For the purposes of this paper we will neglect the influence of
polydispersity and set $V_S \equiv V_S (\bar{r}).$

At finite concentrations, particles start to feel each others presence
and the trajectories are no longer straight.  One observes very
complicated collective effects of the suspension which organizes into
continuously changing regions in
which particles move in different directions.
Denser regions with more particles per unit volume
drag liquid with them and the displaced fluid travels
upward through less dense regions that present less
resistance to the flow. On smaller scales particle velocities change
due to close encounters.  The resulting typical particle trajectories
are highly irregular. Even in the laboratory frame, one finds particles
moving upward in the suspension, describing loops and displaying
settling tendencies only if observed for long enough time.

For illustration we show in Fig.~\ref{fig:trajectories}a
the trajectories of $20$ %
{\it randomly\/} selected particles out of a $1000$ settling in a
suspension at solid volume fraction $\Phi = 0.157$.  Here, the solid
fraction is the dimensionless ratio of the area in the $xy$-plane
covered by disks to the total area in the simulation, i.e.,
\begin{equation} \label{eq:effective_phi}
  \Phi \equiv \sum_i \pi r_i^2/L_xL_y.
\end{equation}
In Fig.~\ref{fig:trajectories}b we
display the same trajectories as viewed from the
comoving frame, i.e., we subtract the mean settling velocity from the
particle velocities in the lab frame and shift the trajectories such
that they all start in one point. For long enough times each
trajectory displays the characteristics of a random walk. Its
direction changes frequently on scales larger than a coherence length,
which measures the average distance that a particle has to travel
before its motion becomes independent of the initial velocity.

In fact, experiments find that the particle velocities at large times
decorrelate~\cite{nicolai}. However, the motion is {\it
 anisotropic,} spreading much more rapidly in the main settling
direction than perpendicular to it.

\subsection{Diffusion constants}
\label{sec:diffusion}


One may describe
the long time behavior of the particle displacements in a sufficiently
large system as a diffusion process in the
frame comoving with the settling particles.
The diffusion is characterized by its mean square displacements
in vertical $y$ and horizontal $x$ direction,
\begin{eqnarray}
  \langle x^2(t)\rangle &\equiv & \frac{1}{N} \sum_i \left[ x_i(t) -
               x_i(0) \right]^2, \\
  \langle y^2(t)\rangle &\equiv & \frac{1}{N} \sum_i \left[ y_i(t) -
               \left( y_i(0) + \langle V\rangle t\right)\right]^2.
\end{eqnarray}

The sum runs over those particles that are above the final bed height
for $t\to\infty$ at the given moment $t$. From their
initial positions we determine the average motion to obtain $\langle
V\rangle.$

As already indicated by Fig.~\ref{fig:trajectories}b, a prominent
feature of hydrodynamic diffusion in batch sedimentation
is the strong anisotropy of the
process with different diffusion constants $D_x$ and $D_y$ in
horizontal and in vertical direction. Both $D_x$ and $D_y$ are
obtained from the long time behavior of the particle displacement by
the relations
\begin{eqnarray} \label{eq:diffusion-constants-1}
    \langle x^2(t) \rangle &\simeq & 2D_xt, \\
    \langle y^2(t) \rangle &\simeq & 2D_yt.\label{eq:diffusion-constants-2}
\end{eqnarray}
In practice, the squared particle displacements saturate due to the
finite system size after passing a regime linear in time. In this
regime, one can determine $D_x$ and $D_y$ using
Eqs.~(\ref{eq:diffusion-constants-1} and
\ref{eq:diffusion-constants-2}). We show typical example for the time
dependence of the mean square displacements in Fig. \ref{fig:diff_t}.
We have normalized both $x$ and $y$-displacements by division through
the squared particle radius, i.e., $\langle x_*^2\rangle \equiv
\langle x^2(t) \rangle / \bar{r}^2$ and accordingly $\langle
y_*^2\rangle\equiv \langle y^2(t) \rangle / \bar{r}^2.$ We obtain the
dimensionless time $t^*$ by division of $t$ through $\bar{r}/V_S$.
The straight lines superposed to the data shall emphasize the
approximately linear behavior for intermediate $t^*.$ The slopes of
these lines, averaged over several runs with different initial
particle positions, are equal to twice the dimensionless diffusion
constants $D^*_x$ and $D^*_y$ which result from division of $D_x$ and
$D_y$ by the quantity $\bar{r}V_S$.

In Fig.~\ref{fig:diffusion}a we plot $D^*_x$ and $D^*_y$ vs. the
solid fraction $\Phi$.
As is observed in real 3D experiments, also here the diffusion constants
increase for small $\Phi$ until in the simulations
a maximum at area fraction $\Phi \approx 0.12$ is
reached. The location of the maximum in the experiments is at
volume fraction $\Phi \approx 0.1$ at a slightly lower value.
At increasingly higher $\Phi$ one observes a renewed
decrease of the diffusion constants. The values of our 2D diffusion
constants lie by a factor of $\approx 2.5\dots 3$ above the 3D
experimental values \cite{nicolai}
or 3D simulation results using Stokesian dynamics
methods \cite{ladd,ladd2}.

Qualitatively, we understand the existence of a maximum at
intermediate volume fractions.  For $\Phi \to 0$, hydrodynamic
interactions between the particles are not very pronounced and the
$D^*$ tend to small values. When $\Phi$ increases, so does the importance of
hydrodynamic interactions between particles and the increasing
complexity of particle trajectories is reflected in an increasing
diffusion constant. When $\Phi$ increases further, then the importance
of close particle encounters also increases, restricting the
motion and consequently reducing the diffusion constants.

The degree of anisotropy of the diffusion is measured by the ratio
$D_y/D_x,$ which is equal to one in the isotropic case.  In the
experiments of Nicolai et al. \cite{nicolai} this ratio takes values
around $5\dots 6$ up to $\Phi \approx 0.20$. For higher solid
fractions the ratio decreases. Such a behavior may be expected when
the importance of particle encounters increases which tends to drive
the diffusion towards isotropy. As shown in Fig.~\ref{fig:diffusion}b
we observe a ratio $D_y/D_x$ of about $4\dots 5$ at small $\Phi$
decreasing to $\approx 2\dots3$ at large $\Phi$.

\subsection{Hindered settling function}
\label{sec:settling}

One of the most easily accessible quantities in a batch settling
experiment is the mean sedimentation velocity $\langle V\rangle$. It
may be obtained by observation of the stable upper sedimentation
front\footnote{
In \protect\cite{Martin94} an initially step-shaped concentration
profile is reported to broaden due to hydrodynamic diffusion
until a stationary smooth profile is reached after $t^* \approx 400$.
}%
, i.e., the interface between batch suspension and clear liquid, or as
average of the velocity of the bulk particles, if the particles
trajectories are traced individually as during an index matched
settling experiment.  The mean sedimentation velocity decreases as the
solid volume fraction $\Phi$ increases. For $\Phi \to 0$ one obtains
$\langle V\rangle = V_S$.

The ratio of sedimentation velocity $\langle V \rangle$ to Stokes
velocity is a dimensionless quantity, known as the hindered settling
function $f(\Phi)$,
\begin{equation}
  f(\Phi) \equiv  \langle V\rangle / V_S.
\end{equation}
The hindered settling function is always smaller than one, decreases
with solid volume fraction and is in most low Reynolds number
$[Re = \rho_{\rm l} V_S r/\eta \ll 1]$ experiments in 3D
well described by the
phenomenological Richardson-Zaki law~\cite{richardson},
\begin{equation}
  f(\Phi) = (1-\Phi)^n,
  \label{eq:richardson-zaki}
\end{equation}
where $n$ typically takes values around $5$ in experiments. Theoretical
considerations by Batchelor et al.~\cite{Batchelor82a} lead to
\begin{equation} \label{eq:batchelor}
f(\Phi) = 1-5.6\Phi
\end{equation}
for $\Phi \to 0$ and $Pe \gg 1$.
Mills et al. \cite{mills} use mean field and
scaling arguments to arrive at the expression
\begin{equation} \label{eq:mills}
 f(\Phi) = (1-\Phi)/[1+k\Phi/(1-\Phi)^3],
\end{equation}
where $k=4.6$ in the non-Brownian regime is a parameter chosen to
match Batchelor's result (\ref{eq:batchelor}). Thus, these expressions all
agree in the low concentration regime and are very well established
experimentally \cite{feuillebois}. Moreover,
Eqs.~(\ref{eq:richardson-zaki}) and (\ref{eq:mills}) are
accurate also at intermediate and high concentrations, not too close
to the random packing threshold.

We display our simulation results in Fig.~\ref{fig:scaled_v}a where
$f(\Phi)$ is plotted vs. the effective 3D volume fraction $\Phi$ of
the simulation.  Clearly, $\langle V\rangle$ decreases when the solid
volume fraction $\Phi$ increases. However, the decrease is not as
sharp as expected from the Richardson-Zaki
law~(\ref{eq:richardson-zaki}) or from Eq.~(\ref{eq:mills}).
Our 2D simulation results are rather fitted approximately by $f(\Phi)
\approx (1-\Phi)^3$ for small $\Phi$. This function is also plotted in
Fig.~\ref{fig:scaled_v}a for comparison.

Apart from the fact that our simulations were done in 2D in contrast to
the 3D experiments, two factors should lead to a sedimentation
velocity that is too high compared to the experiments. (i) There is no
provision in the simulation to take the volume of liquid into account
that is displaced by the particles. Thus, our simulation does not
fully reproduce the net upward backflow effect that otherwise
counteracts the settling.  However, some backflow effect is retained,
because the frictional coupling between liquid and particles
effectively attaches some liquid to every particle. When the particles
settle, this liquid replaces the fluid present in the lower part of
the container and a backflow results \cite{hinch}. (ii) The lattice
spacing sets the scale on which we resolve details of the flow.
Particles act on the liquid by point-like drag forces located on
the four grid points closest to the particle center. The coarser the
grid, the less accurate is this model for the momentum transfer onto
the liquid phase and we underestimate frictional drag because we
underestimate the interstitial liquid velocities.

In fact, reducing the grid spacing leads to a more realistic
representation of the interstitial particle velocities and as a
consequence to smaller sedimentation velocities. We address this
question more systematically in Fig.~\ref{fig:scaled_v}b, where we
plot the simulated values of $f(\Phi),$ for several values of $ \Phi =
0.079,\; 0.16,\; 0.24,\; 0.31\; 0.39$ vs.~the ratio $\Delta
x/\bar{r}.$ In accord with the argumentation above, the simulated
sedimentation velocity slightly decreases for most values as the grid
becomes finer and resolves more details of the interstitial flow
field.


%
%
%
%
%
%

\subsection{Velocity fluctuations}
\label{sec:variances}

Another interesting quantity is the distribution of velocities of the
sedimenting particles. In particular, the standard deviation $\Delta
v_y$ of the $y$-velocity distribution has been measured by Xue et
al.~\cite{xue} and Nicolai et al.~\cite{nicolai} in different
experiments. Here we first focus on the dependence of the velocity
fluctuations on volume fraction at fixed polydispersity $p=0.04,$
corresponding roughly to the particle size dispersion in the
experiment by Nicolai et al.

To calculate the velocity variance $\Delta v_y,$ in
the simulation, we calculate in intervals of about $\Delta t^* =1$ the
standard deviation of the instantaneous $y$-velocity distribution of
selected particles. The selection of particles occurs according to to
the same rules as used for the determination of diffusion constants
(cf. Sec. \ref{sec:diffusion}).  Then we time average the standard
deviations for times $t^* > 20$, which we found sufficient to avoid
significant influence from the initial transient behavior.

We then normalize $\Delta v_y$ by division with $V_S$ to obtain the
dimensionless quantity $\Delta v_y^*.$ Figure~\ref{fig:variance}a
shows $\Delta v_y^*$ as a function of solid fraction $\Phi$. For
comparison, we have added the measured values of
Refs.~\cite{nicolai,xue} into the plot.  The two experimental data
sets agree at large solid fractions, where they show a pronounced
decrease of the fluctuations. At small solid fractions, the data from
\cite{nicolai} passes through a maximum, whereas the data of
\cite{xue} starts at so high $\Phi$ that no such maximum is visible.
Just as the experimental data, also the simulation data shows a
decreasing branch for large solid fractions and displays a maximum of
the fluctuations at intermediate solid fractions, as visible in one of
the experimental data sets.

We explain the existence of a maximum in the velocity fluctuations
by similar factors as have been invoked in
the case of the hydrodynamic diffusion constants, which also
pass through a maximum as functions of $\Phi.$
For increasing volume fraction the fluctuations first increase
due to the more important role of hydrodynamic interactions. For
larger volume fractions, however, the fluctuations are reduced
due to the increasing role of particle encounters, which are
dissipative in nature and tend to reduce fluctuations.

In Fig.~\ref{fig:variance}b we plot the ratio $\Delta v_y^*/\Delta
v_x^*$. For large solid fractions our results match well the
experimental data of Nicolai et al., but do not display the expected
pronounced maximum for lower volume fractions which is visible in the
experimental data. This could be due to the small system size limiting
the fluctuations \cite{Caflish85} and thereby flattening the $\Delta
v_y^*$ curve at intermediate solid fractions.


\section{Conclusions and Outlook}

We have presented a comparatively simple and fast simulation method to
calculate properties of particle suspensions. Although our simulations
have been performed in 2D and involve a set of assumptions that may be
considered to be rather crude, namely the exclusive consideration of
translational degrees of freedom and a simplified local frictional
coupling of the particulate and liquid phase, we have captured the
essential features of 3D sedimentation experiments, like the solid
fraction dependence of (i) the particle self-diffusion constants, (ii)
the velocity fluctuations and (iii) the hindered settling function.
Two advantages of our method are that it is (i) not a priori limited
to 2D or (ii) to small Reynolds numbers. Moreover, more
sophistication may enter the drag expression in order to improve the
quantitative agreement with experiments once the program has been
extended to 3D.  We are currently working on the extension of the
method both to 3D and to Reynolds numbers around $100.$

\section*{Acknowledgments}

We are grateful for lively and helpful discussions with F.
Feuillebois, J. Gallas, E. Guazzelli, J. Hinch, J. Hofhaus, L.
Kjeldgaard, H.-G. Mattutis, S. Melin, C. Moukarzel, T. P\"oschel,
H.-J. Tillemans, W. Verm\"ohlen, and M. Weimer and with several
participants of the NATO ASI ``Mobile Particulate Systems.'' GR thanks
the statistical physics group at the University of Marburg for
fruitful discussions and comments. We are particularly grateful to M.
Rutgers~\cite{xue} and H. Nicolai~\cite{nicolai} who supplied
experimental data for comparison.

The numerical computations have been performed on workstations at the
HLRZ, on the 144-node Intel XP/S 10 Paragon and the SP1 at the ZAM,
Forschungszentrum J\"ulich and on the SP1 of the Gesellschaft f\"ur
Mathematik und Datenverarbeitung in Bonn.


\newpage

\section*{Figure Captions}
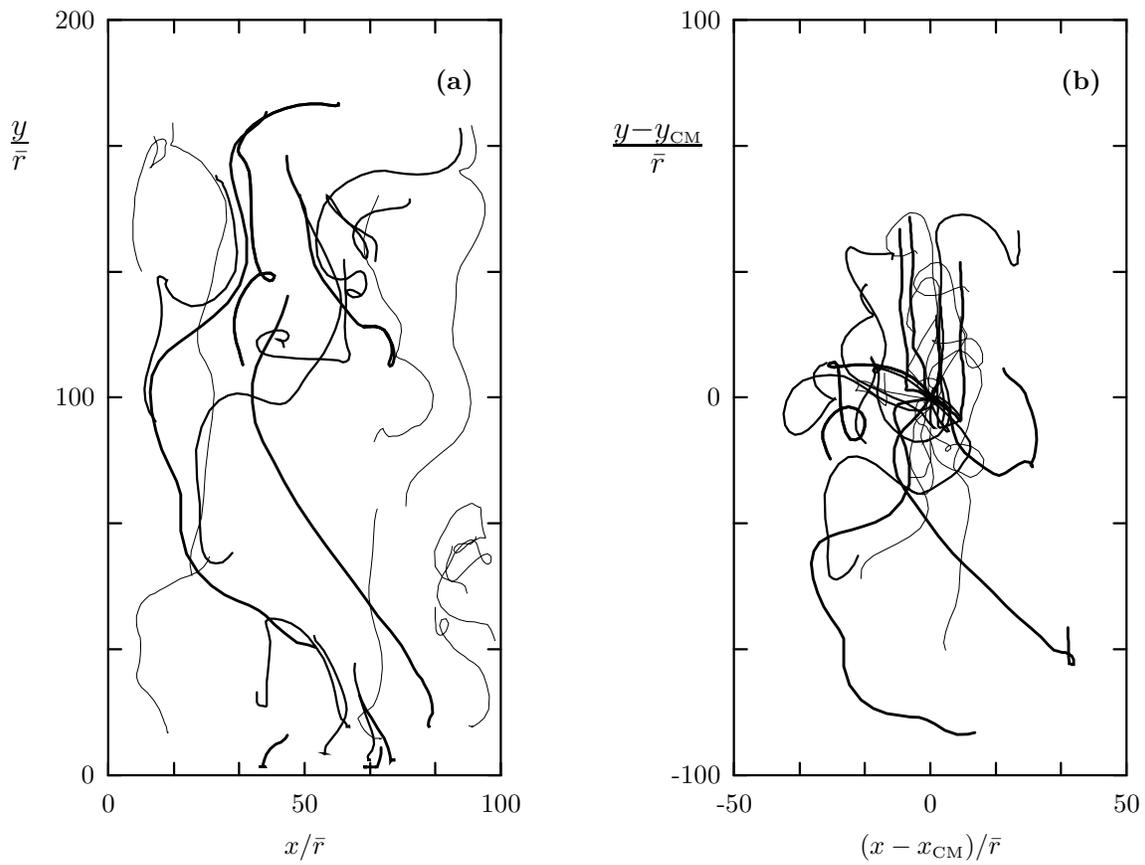
\begin{figure}[p]
  \hbox{\input trace_1 \hfill \input trace_2 }
  \caption{}
  \label{fig:trajectories}
\end{figure}

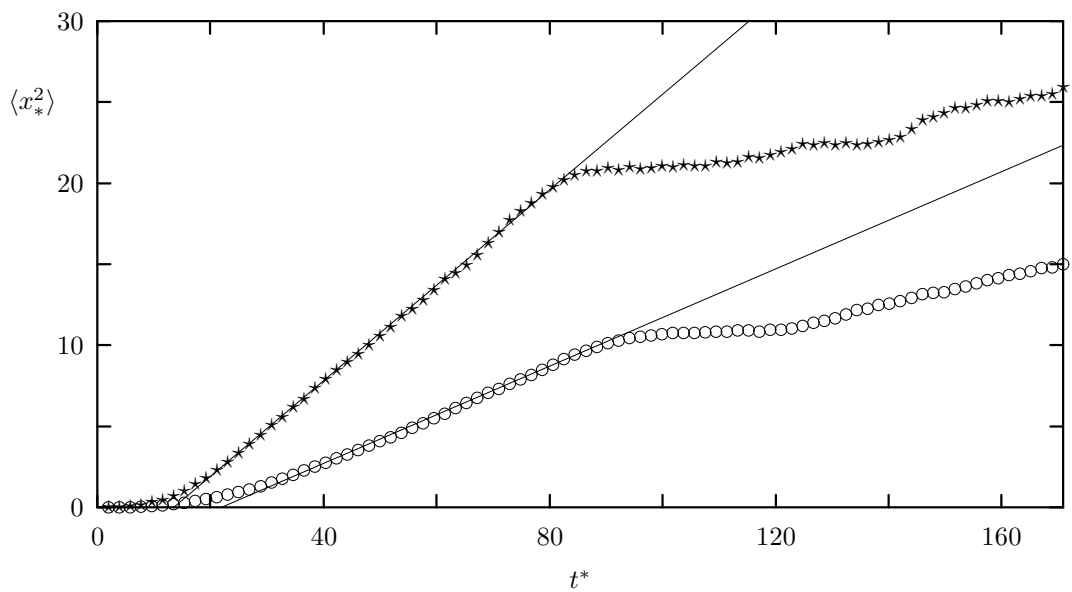
\begin{figure}[p]
   \centerline{\input diff  
              }
  \caption{}
  \label{fig:diff_t}
\end{figure}

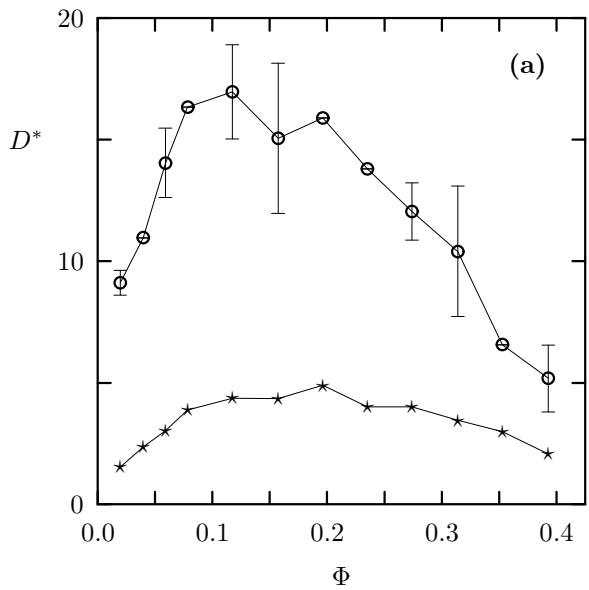
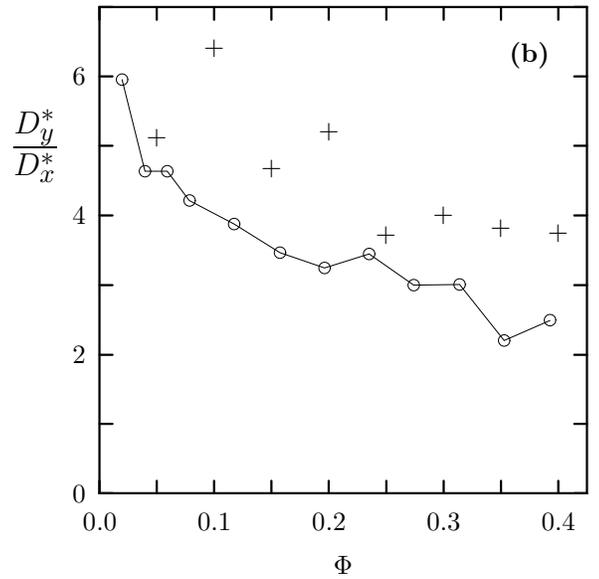
\begin{figure}[p]
  \hbox{ \input sed_new1b  
  \hfill \input sed_new1c 
        }
  \caption{}
  \label{fig:diffusion}
\end{figure}

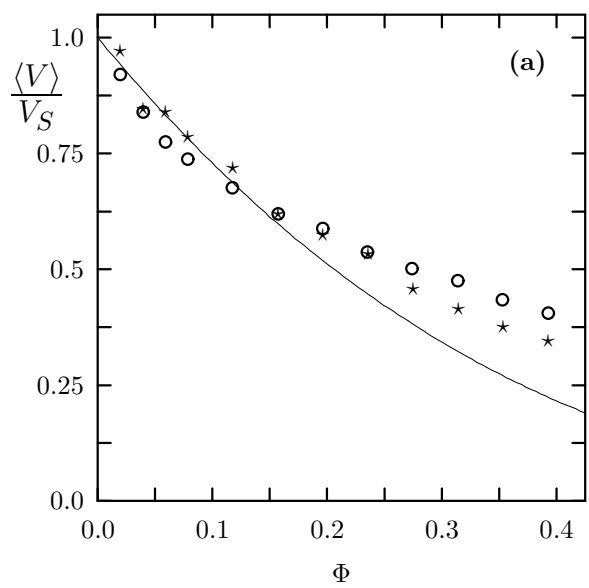
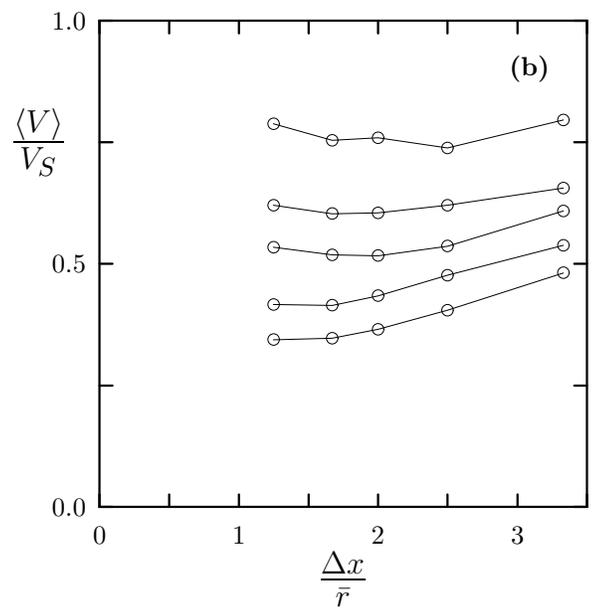
\begin{figure}[p]
  \hbox{ \input sed_new3a   
  \hfill \input sed_new3d  
       }
  \caption{}
  \label{fig:scaled_v}
\end{figure}

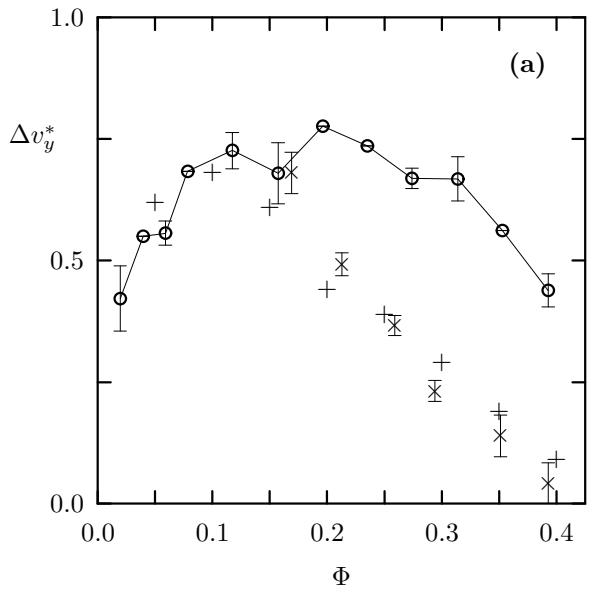
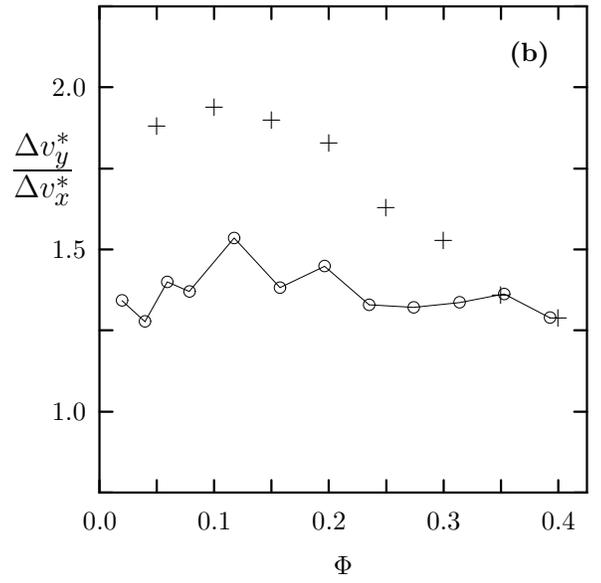
\begin{figure}[p]
  \hbox{ \input sed_new4a     
  \hfill \input sed_new4b     
        }
  \caption{}
  \label{fig:variance}
\end{figure}

\begin{description}
\item[\bf Figure~\ref{fig:trajectories}:] Sedimentation of
  polydisperse spheres at $\Phi=0.157$: {\bf (a)}
  particle trajectories in the laboratory frame,
  {\bf (b)} particle trajectories in the comoving frame of the
  sedimenting cloud. The particle coordinates have been shifted to
  start at a common origin and the coordinate change due to the
  mean settling velocity has been subtracted.

\item[\bf Figure~\ref{fig:diff_t}:] Dimensionless $x$ ($\bigcirc$)
  and $y$ ($\star$) mean square displacements [$\langle
  x^2_*\rangle\equiv \langle x^2\rangle / \bar{r}^2$ and $\langle
  y^2_*\rangle\equiv \langle y^2\rangle / \bar{r}^2$] as functions of
  dimensionless time $t^* \equiv t V_S/\bar{r}$ averaged over four
  runs. The two solid lines indicate a linear approximation to
  the data in an intermediate time regime. The slope of these lines
  equals twice the dimensionless diffusion constants $D_x^*$ and
  $D_y^*.$

\item[\bf Figure~\ref{fig:diffusion}:] {\bf (a)} Dimensionless
  particle self-diffusion constants $D^*_x$ ($\star$) and $D^*_y$
  ($\bigcirc$) as a function of solid fraction $\Phi.$ We have
  displayed errorbars on selected data points to indicate deviations
  of the mean obtained from $n$ averages over different initial
  configurations [$\Phi=0.16$: $n=6$, $\Phi=0.31$: $n=5$, all other
  concentrations: $n=3$].  {\bf (b)} Dimensionless ratio $D_y/D_x$ of
  self-diffusivity parallel ($D_y$) and perpendicular ($D_x$) to
  settling direction ($\bigcirc$) and the experimental data from
  Nicolai et al. ~\protect\cite{nicolai} $(+)$.

\item[\bf Figure~\ref{fig:scaled_v}:] {\bf (a)} Hindered settling
  function $f(\Phi)$.
  The graph corresponds to data with a fixed ratio of grid spacing to
  particle size $\Delta x/\bar{r} = 2.5\; (\bigcirc)$ and $ 1.25\;
  (\star)$ respectively. The solid curve is given by $(1-\Phi)^3$.
  {\bf (b)} To demonstrate the influence of grid spacing on the
  accuracy of our results, we display the hindered settling function
  for different solid fraction $\Phi$ as a function of the ratio of
  grid spacing to average particle radius $\Delta x/\bar{r}$.  From
  top to bottom the curves correspond to $ \Phi = 0.079,\; 0.16,\;
  0.24,\; 0.31$ and $0.39$ respectively.
\item[\bf Figure~\ref{fig:variance}:] {\bf (a)} Dependence of the
  width of the velocity distribution $\Delta v_y^*$ on the volume
  fraction. We show simulation data for a polydispersity of $p=0.04$
  $(\bigcirc)$, measured data by Nicolai et al. for $p\approx 0.04$
  $(+)$ and Xue et al.~\protect\cite{xue} for $p=0.01$ $(\times).$ In
  all cases, the fluctuations $\Delta v_y^*$ and $\Delta v^*_x$ have
  been made non-dimensional by division through the Stokes velocity.
  Errorbars on selected data points indicate the deviation of the mean
  for averages over different initial conditions, their number being
  equal to the numbers in Fig.  \protect\ref{fig:diffusion}a.
  {\bf (b)} Ratio of $\Delta v_y^*$ to $\Delta v_x^*$. We show simulated
  data at $p=0.04$ $(\bigcirc)$ and the experimental results of
  Nicolai et al.~\protect\cite{nicolai} $(+)$.


\end{description}

\newpage

\begin{table}[h]
\begin{tabular}{clrl}
 $L_x$          & system width            & $3$     & $cm$ \\
 $L_y$          & system height           & $6$     & $cm$ \\
 $\eta$         & liquid shear viscosity  & $0.192$ & $g/cm\,s$ \\
 $\rho_{\rm l}$ & liquid density          & $1.09$  & $g/cm^3$  \\
 $z$            & Stokes force reference length & $0.06$ & $cm$ \\
 $\rho_{\rm p}$ & particle density        & $2.53$  & $g/cm^3$  \\
 $\bar{r}$      & average particle radius & $0.03$ & $cm$ \\
 $p$            & polydispersity          & $0.04$ & \\
 $k_n$          & repulsion constant        & $1\times 10^6$ & $g/s^2$ \\
 $g$            & gravitational acceleration & $980$ & $cm/s^2$ \\
\end{tabular}
\caption{List of fixed simulation parameters. Unless otherwise
  specified, simulations runs have been performed
  using the parameters in this table.}
\label{tab:1}
\end{table}


\newpage

\end{document}

%% file: trace_1
\setlength{\unitlength}{0.240900pt}
\begin{picture}(900,1323)(0,0)
\tenrm
\thicklines \path(220,113)(240,113)
\thicklines \path(836,113)(816,113)
\put(198,113){\makebox(0,0)[r]{0}}
\thicklines \path(220,311)(240,311)
\thicklines \path(836,311)(816,311)
\thicklines \path(220,509)(240,509)
\thicklines \path(836,509)(816,509)
\thicklines \path(220,707)(240,707)
\thicklines \path(836,707)(816,707)
\put(198,707){\makebox(0,0)[r]{100}}
\thicklines \path(220,904)(240,904)
\thicklines \path(836,904)(816,904)
\thicklines \path(220,1102)(240,1102)
\thicklines \path(836,1102)(816,1102)
\thicklines \path(220,1300)(240,1300)
\thicklines \path(836,1300)(816,1300)
\put(198,1300){\makebox(0,0)[r]{200}}
\thicklines \path(220,113)(220,133)
\thicklines \path(220,1300)(220,1280)
\put(220,68){\makebox(0,0){0}}
\thicklines \path(323,113)(323,133)
\thicklines \path(323,1300)(323,1280)
\thicklines \path(425,113)(425,133)
\thicklines \path(425,1300)(425,1280)
\thicklines \path(528,113)(528,133)
\thicklines \path(528,1300)(528,1280)
\put(528,68){\makebox(0,0){50}}
\thicklines \path(631,113)(631,133)
\thicklines \path(631,1300)(631,1280)
\thicklines \path(733,113)(733,133)
\thicklines \path(733,1300)(733,1280)
\thicklines \path(836,113)(836,133)
\thicklines \path(836,1300)(836,1280)
\put(836,68){\makebox(0,0){100}}
\thicklines \path(220,113)(836,113)(836,1300)(220,1300)(220,113)
\put(80,1102){\makebox(0,0){\Large $\frac{y}{\bar{r}}$}}
\put(528,0){\makebox(0,0){$x/\bar{r}$}}
\put(733,1201){\makebox(0,0)[l]{\bf (a)}}
\thinlines \path(760,473)(760,473)(758,463)(754,453)(751,444)(750,433)(753,422)(759,418)(766,420)(775,430)(786,441)(798,452)(808,459)(815,461)(820,458)(821,453)(820,448)(816,445)(812,446)(808,450)(805,457)(802,464)(800,472)(798,478)(794,482)(789,483)(783,481)(776,476)(769,468)(763,461)(758,450)(752,439)(745,428)(738,411)(736,394)(742,378)(755,365)(771,358)(783,355)(795,354)(805,352)(815,350)(820,343)(824,334)(827,325)
\thicklines \path(521,1025)(521,1025)(522,1016)(525,1005)(530,991)(537,978)(545,961)(554,945)(563,926)(571,909)(578,891)(584,872)(586,853)(583,833)(576,813)(568,794)(558,776)(547,760)(535,745)(525,731)(518,719)(511,710)(505,703)(498,698)(489,696)(474,698)(458,704)(439,711)(422,713)(408,710)(394,702)(379,686)(366,660)(358,624)(359,580)(364,539)(364,507)(365,482)(368,465)(373,455)(380,448)(389,446)(399,448)(408,455)(415,463)
\Thicklines \path(501,176)(501,176)(498,171)(491,165)(483,162)(475,155)(468,146)(465,138)(463,130)(465,127)(464,127)(465,127)(466,127)(466,127)(467,127)(467,127)(467,127)(465,127)(465,127)(464,127)(462,127)(461,127)(460,127)(459,127)(458,127)(457,127)(457,127)(457,127)(458,127)(458,127)(459,127)(459,127)(459,127)(458,127)(458,127)(459,127)(459,127)(459,127)(459,127)(459,127)(459,127)(458,127)(459,127)(459,127)(459,127)
\thinlines \path(644,1024)(644,1024)(642,1019)(639,1012)(634,1001)(629,989)(625,975)(621,956)(622,933)(628,916)(636,906)(640,895)(644,880)(650,863)(661,848)(672,835)(679,822)(680,810)(678,801)(677,792)(675,782)(672,776)(672,771)(677,768)(686,765)(698,757)(711,746)(723,731)(730,711)(728,690)(719,673)(709,663)(701,661)(695,658)(690,654)(682,649)(674,646)(667,646)(660,645)(655,650)(652,655)(651,658)(647,656)(642,649)(637,637)
\thicklines \path(453,244)(453,244)(453,235)(453,228)(455,224)(458,223)(463,222)(467,221)(469,223)(469,228)(469,237)(469,247)(469,260)(470,273)(472,289)(473,306)(474,323)(471,338)(468,348)(467,355)(472,359)(481,359)(493,357)(507,352)(521,343)(534,330)(546,314)(558,296)(566,277)(572,259)(576,242)(579,225)(582,212)(586,202)(590,198)(594,193)(594,189)(595,189)(597,190)(598,189)(598,189)(598,189)(598,189)(598,190)(598,191)
\Thicklines \path(614,239)(614,239)(617,231)(624,217)(633,203)(643,186)(653,171)(659,156)(662,145)(663,138)(663,136)(664,137)(666,137)(668,137)(670,137)(669,137)(670,137)(670,137)(670,137)(670,137)(669,137)(667,137)(666,137)(665,137)(665,137)(665,137)(664,137)(664,137)(665,137)(666,137)(667,137)(667,137)(667,137)(667,137)(667,137)(667,137)(667,137)(667,137)(667,137)(667,137)(667,137)(667,137)(667,137)(667,137)(667,137)
\thinlines \path(379,457)(379,457)(376,449)(369,440)(360,433)(346,426)(333,422)(320,415)(310,409)(301,404)(293,399)(285,396)(278,390)(271,382)(267,373)(264,363)(264,354)(265,347)(267,337)(270,329)(271,321)(271,311)(270,302)(269,292)(267,283)(265,274)(265,266)(267,259)(272,251)(281,243)(291,231)(300,217)(305,204)(308,196)(311,189)(312,182)(313,181)(314,180)(314,180)(313,181)(313,181)(313,181)(313,181)(313,181)(313,181)
\thicklines \path(772,1129)(772,1129)(774,1120)(774,1105)(771,1090)(764,1078)(753,1068)(738,1061)(722,1058)(705,1057)(687,1058)(671,1060)(654,1063)(637,1062)(617,1056)(595,1045)(575,1030)(559,1009)(550,982)(546,954)(548,929)(555,911)(564,899)(574,894)(584,892)(594,893)(603,899)(610,904)(616,905)(621,902)(623,898)(624,892)(625,885)(624,878)(621,872)(617,867)(612,865)(604,864)(597,868)(594,873)(595,875)(599,876)(604,875)(610,872)(615,869)
\Thicklines \path(502,866)(502,866)(499,855)(493,837)(483,815)(472,795)(461,775)(452,755)(446,729)(446,699)(451,667)(463,637)(479,607)(496,575)(517,539)(540,504)(562,474)(583,446)(604,418)(624,390)(643,363)(663,340)(680,317)(692,296)(699,282)(705,272)(712,264)(716,256)(721,247)(725,238)(728,225)(729,211)(725,204)(723,199)(722,192)(723,190)(723,190)(723,190)(723,190)(723,190)(723,190)(723,190)(723,190)(724,190)(724,190)
\thinlines \path(292,1117)(292,1117)(293,1113)(294,1110)(296,1106)(297,1099)(296,1091)(293,1084)(291,1078)(290,1074)(288,1071)(287,1068)(286,1068)(288,1070)(291,1073)(297,1077)(303,1081)(306,1084)(308,1086)(309,1090)(309,1096)(309,1101)(311,1106)(310,1111)(305,1113)(297,1111)(288,1103)(281,1090)(276,1075)(272,1062)(268,1049)(265,1035)(262,1022)(261,1009)(260,997)(259,985)(259,973)(259,963)(259,954)(260,946)(261,939)(263,931)(265,924)(268,915)(272,905)
\thicklines \path(543,332)(543,332)(545,325)(551,318)(558,309)(566,295)(573,281)(580,266)(586,251)(590,238)(593,226)(594,216)(592,205)(587,194)(580,182)(570,170)(562,159)(560,153)(560,148)(559,148)(559,148)(560,147)(559,147)(558,147)(558,147)(558,147)(558,147)(558,147)(558,147)(559,147)(559,147)(560,147)(559,147)(559,147)(559,146)(559,146)(559,146)(559,147)(559,147)(559,147)(560,147)(559,147)(559,147)(560,147)(560,147)
\Thicklines \path(468,1155)(468,1155)(465,1147)(455,1137)(442,1126)(429,1116)(419,1104)(414,1089)(412,1073)(414,1052)(421,1025)(430,994)(437,957)(437,918)(426,883)(407,855)(385,834)(366,819)(349,806)(331,793)(313,779)(299,763)(289,741)(286,715)(288,684)(295,654)(304,625)(316,599)(327,575)(334,554)(334,529)(334,497)(343,461)(361,432)(382,411)(403,397)(423,387)(441,380)(456,372)(468,363)(481,350)(497,336)(516,322)(533,318)(545,314)
\thinlines \path(790,1134)(790,1134)(792,1125)(795,1113)(797,1102)(797,1092)(796,1084)(794,1078)(790,1072)(785,1067)(779,1066)(774,1064)(771,1063)(773,1058)(780,1053)(789,1048)(798,1038)(806,1027)(811,1011)(811,994)(807,977)(800,960)(791,943)(782,928)(773,913)(768,898)(766,883)(766,869)(765,854)(763,837)(760,814)(762,787)(772,764)(781,741)(787,720)(789,698)(786,676)(777,654)(762,635)(742,619)(723,603)(706,588)(694,573)(688,556)(687,536)
\thicklines \path(693,1018)(693,1018)(693,1013)(688,1008)(681,1005)(671,1004)(661,1003)(650,1001)(638,999)(627,994)(617,988)(611,981)(606,972)(604,960)(607,948)(612,940)(617,934)(619,930)(621,927)(624,925)(626,927)(627,936)(623,949)(614,963)(602,976)(589,988)(577,999)(568,1010)(563,1019)(563,1024)(566,1022)(570,1018)(575,1010)(581,1000)(587,990)(595,982)(605,975)(616,969)(626,962)(634,958)(639,952)(641,946)(641,940)(640,931)(639,921)
\Thicklines \path(581,1169)(581,1169)(581,1166)(577,1165)(567,1166)(554,1168)(537,1168)(518,1166)(498,1161)(479,1154)(462,1145)(448,1135)(438,1126)(432,1116)(428,1108)(425,1102)(424,1094)(426,1088)(431,1082)(436,1075)(441,1061)(443,1041)(444,1011)(445,977)(447,945)(452,920)(459,902)(466,894)(473,890)(478,891)(480,895)(481,898)(475,902)(467,902)(455,899)(445,890)(435,877)(427,862)(421,845)(418,829)(418,812)(421,794)(425,780)(428,768)(431,758)
\thinlines \path(321,1138)(321,1138)(321,1132)(322,1124)(322,1116)(321,1109)(319,1106)(318,1104)(318,1103)(319,1103)(321,1102)(326,1101)(332,1098)(338,1094)(346,1091)(353,1086)(359,1082)(366,1077)(372,1069)(382,1057)(393,1039)(404,1015)(406,985)(399,958)(392,938)(387,914)(383,889)(378,869)(371,848)(365,830)(360,812)(360,793)(364,777)(373,762)(382,743)(386,713)(383,675)(377,634)(372,590)(370,550)(366,518)(358,491)(350,466)(348,444)(351,427)
\thicklines \path(590,923)(590,923)(590,914)(591,903)(592,894)(592,884)(592,875)(593,865)(593,852)(592,837)(591,819)(592,800)(595,786)(599,775)(597,766)(593,763)(588,762)(581,765)(568,766)(553,767)(535,767)(515,768)(496,770)(480,771)(467,774)(458,779)(454,788)(453,796)(459,803)(469,808)(482,812)(493,812)(500,809)(503,804)(502,799)(498,796)(493,795)(489,797)(486,798)(485,798)(487,797)(489,797)(492,796)(494,792)(496,786)
\Thicklines \path(500,1086)(500,1086)(500,1080)(501,1074)(502,1065)(505,1056)(509,1044)(516,1030)(523,1017)(528,1004)(531,987)(534,965)(537,941)(544,915)(557,891)(573,870)(589,850)(602,833)(612,822)(622,817)(633,818)(644,817)(653,813)(659,805)(663,795)(665,784)(665,776)(663,768)(661,763)(660,759)(662,756)(665,757)(667,761)(669,770)(668,780)(666,789)(662,797)(657,805)(653,812)(647,816)(639,819)(631,818)(625,819)(622,819)(621,817)
\thinlines \path(797,488)(797,488)(796,483)(792,479)(788,476)(782,477)(776,477)(769,476)(764,472)(762,467)(765,462)(771,464)(778,470)(786,478)(796,483)(804,487)(809,486)(812,486)(812,487)(811,490)(809,495)(807,505)(806,514)(805,523)(805,530)(804,534)(801,535)(799,536)(795,538)(792,540)(789,540)(784,538)(777,535)(767,529)(757,516)(750,502)(741,486)(732,465)(733,441)(745,424)(759,418)(769,416)(775,414)(780,410)(783,399)
\thicklines \path(389,1055)(389,1055)(391,1049)(395,1047)(398,1045)(400,1043)(400,1041)(401,1038)(404,1032)(409,1019)(413,1002)(417,982)(419,960)(422,935)(419,905)(404,878)(386,861)(368,852)(350,850)(335,856)(322,864)(315,873)(311,880)(310,885)(311,889)(312,891)(310,892)(306,895)(301,897)(298,895)(298,887)(300,876)(302,864)(304,850)(304,835)(302,820)(299,805)(294,788)(289,773)(284,756)(279,738)(277,720)(278,702)(284,684)(294,669)
\Thicklines \path(648,157)(648,157)(648,147)(646,138)(645,135)(642,126)(642,126)(640,126)(639,126)(640,126)(639,126)(638,126)(633,126)(627,126)(623,126)(623,126)(625,126)(626,126)(626,126)(625,126)(624,126)(623,126)(622,126)(621,126)(621,126)(620,126)(620,126)(620,126)(620,126)(621,126)(622,126)(622,126)(623,126)(623,126)(623,126)(622,126)(622,126)(622,126)(622,126)(622,126)(622,126)(622,126)(622,126)(622,126)(622,126)
\thinlines \path(733,377)(733,377)(733,369)(733,359)(734,350)(734,343)(736,336)(740,333)(744,334)(747,336)(750,340)(752,344)(754,352)(750,359)(743,358)(740,349)(744,337)(750,328)(756,321)(762,313)(770,306)(779,300)(786,294)(793,290)(802,283)(809,270)(814,255)(818,237)(817,217)(810,203)(801,195)(795,191)(792,190)(791,190)(790,190)(791,190)(791,190)(792,190)(792,190)(792,190)(789,190)(786,191)(786,191)(786,191)(786,191)
\thicklines \path(607,289)(607,289)(606,278)(608,262)(613,244)(618,226)(624,208)(629,194)(633,184)(636,173)(637,162)(634,152)(630,147)(628,145)(627,142)(629,137)(631,137)(632,137)(632,137)(632,137)(631,137)(629,137)(628,137)(627,137)(627,137)(627,137)(626,137)(626,137)(627,137)(628,137)(628,137)(629,137)(629,137)(629,137)(629,137)(629,137)(629,137)(629,137)(628,137)(628,137)(629,137)(628,137)(629,137)(629,137)(629,137)
\Thicklines \path(581,1169)(581,1169)(581,1166)(577,1165)(567,1166)(554,1168)(537,1168)(518,1166)(498,1161)(479,1154)(462,1145)(448,1135)(438,1126)(432,1116)(428,1108)(425,1102)(424,1094)(426,1088)(431,1082)(436,1075)(441,1061)(443,1041)(444,1011)(445,977)(447,945)(452,920)(459,902)(466,894)(473,890)(478,891)(480,895)(481,898)(475,902)(467,902)(455,899)(445,890)(435,877)(427,862)(421,845)(418,829)(418,812)(421,794)(425,780)(428,768)(431,758)
\thinlines \path(642,532)(642,532)(640,525)(638,513)(637,500)(636,484)(635,465)(633,445)(631,427)(629,410)(629,394)(631,379)(634,363)(639,343)(645,324)(649,303)(650,281)(648,261)(645,245)(641,233)(636,223)(629,217)(622,213)(616,207)(613,204)(612,203)(612,201)(614,198)(617,191)(622,184)(628,178)(634,173)(637,170)(639,171)(641,171)(646,171)(646,171)(646,171)(646,170)(647,170)(647,170)(647,170)(648,171)(648,171)(648,170)
\end{picture}

%% file: trace_2
\setlength{\unitlength}{0.240900pt}
\begin{picture}(900,1323)(-50,0)
\tenrm
\thicklines \path(220,113)(240,113)
\thicklines \path(836,113)(816,113)
\put(198,113){\makebox(0,0)[r]{-100}}
\thicklines \path(220,311)(240,311)
\thicklines \path(836,311)(816,311)
\thicklines \path(220,509)(240,509)
\thicklines \path(836,509)(816,509)
\thicklines \path(220,707)(240,707)
\thicklines \path(836,707)(816,707)
\put(198,707){\makebox(0,0)[r]{0}}
\thicklines \path(220,904)(240,904)
\thicklines \path(836,904)(816,904)
\thicklines \path(220,1102)(240,1102)
\thicklines \path(836,1102)(816,1102)
\thicklines \path(220,1300)(240,1300)
\thicklines \path(836,1300)(816,1300)
\put(198,1300){\makebox(0,0)[r]{100}}
\thicklines \path(220,113)(220,133)
\thicklines \path(220,1300)(220,1280)
\put(220,68){\makebox(0,0){-50}}
\thicklines \path(323,113)(323,133)
\thicklines \path(323,1300)(323,1280)
\thicklines \path(425,113)(425,133)
\thicklines \path(425,1300)(425,1280)
\thicklines \path(528,113)(528,133)
\thicklines \path(528,1300)(528,1280)
\put(528,68){\makebox(0,0){0}}
\thicklines \path(631,113)(631,133)
\thicklines \path(631,1300)(631,1280)
\thicklines \path(733,113)(733,133)
\thicklines \path(733,1300)(733,1280)
\thicklines \path(836,113)(836,133)
\thicklines \path(836,1300)(836,1280)
\put(836,68){\makebox(0,0){50}}
\thicklines \path(220,113)(836,113)(836,1300)(220,1300)(220,113)
\put(100,1102){\makebox(0,0){\Large $\frac{y-y_{\mbox{\tiny CM}}}{\bar{r}}$}}
\put(528,0){\makebox(0,0){$(x-x_{\mbox{\tiny CM}})/\bar{r}$}}
\put(733,1201){\makebox(0,0)[l]{\bf (b)}}
\thinlines \path(528,707)(528,707)(526,704)(522,702)(519,701)(518,699)(520,697)(526,701)(533,712)(542,730)(553,750)(565,769)(575,785)(581,795)(585,801)(586,805)(584,808)(581,813)(576,822)(572,834)(568,848)(566,862)(564,878)(561,891)(558,902)(553,911)(547,916)(541,919)(535,918)(529,917)(524,913)(518,909)(511,904)(504,894)(502,884)(507,874)(520,867)(535,867)(546,869)(558,874)(568,878)(577,881)(582,880)(585,876)(589,873)
\thicklines \path(528,707)(528,707)(528,704)(531,701)(536,696)(543,691)(551,683)(560,675)(569,665)(577,657)(584,646)(590,637)(591,627)(588,615)(580,604)(572,592)(561,583)(550,575)(538,568)(528,562)(520,558)(514,556)(508,556)(500,559)(492,564)(478,573)(462,586)(443,601)(427,610)(413,614)(399,613)(384,602)(371,583)(362,554)(363,517)(368,482)(367,456)(368,438)(370,427)(375,422)(381,421)(390,425)(400,433)(409,445)(415,458)
\Thicklines \path(528,707)(528,707)(525,708)(519,711)(510,716)(502,718)(495,718)(491,718)(490,719)(491,724)(490,732)(491,741)(492,749)(492,758)(492,766)(492,775)(491,783)(489,791)(488,799)(487,807)(485,815)(484,822)(483,830)(482,837)(481,844)(481,852)(481,859)(482,866)(483,873)(484,880)(484,886)(484,893)(484,900)(484,906)(483,913)(483,919)(483,926)(482,932)(482,938)(481,944)(481,949)(480,955)(480,961)(480,966)(480,971)
\thinlines \path(528,707)(528,707)(526,707)(522,709)(518,707)(512,703)(508,698)(505,687)(505,673)(511,665)(519,663)(523,660)(526,654)(532,646)(543,639)(554,634)(560,630)(560,626)(558,625)(557,624)(555,622)(552,623)(552,626)(556,630)(566,633)(578,633)(592,630)(604,622)(612,609)(610,594)(601,585)(591,581)(583,585)(577,589)(571,591)(563,593)(555,597)(547,602)(540,608)(534,618)(531,630)(529,638)(525,641)(520,639)(515,633)
\thicklines \path(528,707)(528,707)(528,705)(528,706)(529,710)(533,718)(538,726)(541,733)(543,744)(543,757)(543,775)(542,794)(542,815)(543,837)(544,861)(545,886)(545,911)(542,934)(538,952)(538,967)(542,979)(551,987)(563,992)(578,994)(591,993)(605,988)(618,979)(630,968)(638,955)(645,944)(649,934)(652,924)(655,917)(659,914)(663,916)(665,917)(665,920)(665,927)(667,933)(668,938)(667,944)(667,949)(667,955)(667,961)(666,968)
\Thicklines \path(528,707)(528,707)(531,706)(538,700)(547,694)(557,686)(567,680)(572,673)(575,670)(576,671)(576,679)(577,688)(579,696)(580,705)(581,713)(581,722)(581,730)(581,738)(580,746)(580,754)(579,762)(577,769)(576,777)(575,784)(575,792)(575,799)(575,806)(576,814)(577,820)(578,827)(579,834)(579,841)(579,847)(579,854)(579,860)(578,867)(578,873)(577,879)(576,886)(576,891)(576,897)(576,903)(575,908)(575,914)(575,919)
\thinlines \path(528,707)(528,707)(525,706)(519,705)(509,707)(496,709)(482,712)(469,714)(459,716)(450,720)(442,723)(434,729)(426,732)(419,732)(414,732)(411,730)(410,729)(411,730)(413,729)(415,729)(416,728)(416,726)(415,724)(414,722)(413,720)(411,718)(411,717)(414,718)(419,717)(428,716)(439,710)(448,703)(453,697)(456,695)(458,694)(458,694)(459,699)(460,704)(459,711)(457,717)(457,723)(457,728)(456,734)(456,739)(456,745)
\thicklines \path(528,707)(528,707)(530,704)(530,697)(527,691)(519,688)(509,686)(494,687)(477,693)(460,700)(442,710)(426,721)(409,732)(392,740)(371,742)(349,740)(328,733)(312,720)(302,701)(298,681)(300,663)(307,653)(316,648)(326,650)(336,656)(346,665)(356,677)(364,689)(370,698)(375,702)(377,704)(378,705)(379,704)(378,704)(375,704)(371,706)(364,710)(356,716)(348,725)(345,736)(346,745)(350,751)(354,756)(360,758)(365,761)
\Thicklines \path(528,707)(528,707)(525,703)(519,693)(509,680)(498,668)(487,657)(478,645)(472,627)(471,606)(476,583)(488,562)(504,540)(521,516)(541,489)(563,462)(585,440)(606,421)(626,401)(646,380)(665,362)(685,345)(702,330)(714,317)(722,310)(728,308)(735,306)(740,306)(745,304)(750,301)(753,295)(753,287)(750,287)(747,289)(746,289)(746,292)(745,299)(745,305)(745,312)(745,317)(745,323)(744,329)(744,334)(744,340)(744,345)
\thinlines \path(528,707)(528,707)(529,710)(530,715)(532,720)(533,722)(531,722)(529,723)(527,725)(525,730)(524,735)(522,741)(521,750)(522,760)(525,771)(531,784)(536,796)(539,808)(540,818)(541,830)(540,843)(541,856)(542,868)(542,880)(537,890)(530,895)(521,894)(514,888)(510,881)(506,874)(502,868)(499,861)(497,854)(495,848)(494,842)(493,836)(492,831)(492,827)(491,825)(491,822)(492,821)(493,818)(495,817)(498,813)(502,809)
\thicklines \path(528,707)(528,707)(530,706)(535,708)(542,707)(550,702)(558,696)(565,689)(570,683)(574,678)(577,675)(577,673)(575,672)(570,668)(562,665)(552,661)(543,659)(541,661)(540,664)(540,672)(539,679)(540,686)(540,693)(539,701)(539,708)(539,715)(539,723)(540,730)(541,737)(541,743)(542,750)(542,757)(542,763)(542,770)(542,776)(541,783)(540,789)(540,795)(539,802)(539,808)(539,813)(538,819)(538,824)(538,830)(538,835)
\Thicklines \path(528,707)(528,707)(525,705)(515,703)(502,701)(489,699)(478,696)(473,690)(471,682)(473,670)(480,652)(489,629)(495,600)(495,570)(483,544)(464,523)(442,511)(422,504)(404,499)(386,494)(369,488)(354,479)(345,465)(341,446)(344,422)(351,400)(361,378)(373,359)(385,342)(392,328)(392,309)(392,284)(401,255)(419,232)(439,218)(460,210)(480,207)(497,205)(511,204)(523,200)(535,194)(551,184)(570,177)(587,178)(598,180)
\thinlines \path(528,707)(528,707)(530,703)(534,700)(536,698)(535,697)(534,697)(532,699)(528,702)(523,705)(516,712)(511,719)(508,727)(510,731)(516,734)(524,737)(533,736)(541,732)(545,725)(545,716)(541,707)(534,696)(525,687)(516,680)(508,672)(503,664)(502,657)(502,649)(502,642)(499,632)(496,615)(499,595)(508,578)(517,562)(523,547)(525,532)(521,516)(512,501)(496,488)(476,477)(456,467)(439,458)(426,448)(420,437)(419,422)
\thicklines \path(528,707)(528,707)(527,708)(523,712)(516,717)(506,725)(495,733)(484,740)(472,745)(461,749)(451,752)(445,753)(440,753)(438,749)(440,746)(444,746)(448,749)(450,753)(452,758)(454,764)(457,773)(458,790)(454,810)(445,831)(433,852)(420,871)(409,889)(400,908)(396,924)(396,935)(399,940)(403,943)(408,941)(413,938)(419,935)(427,933)(436,932)(447,932)(456,932)(464,933)(468,933)(470,932)(470,932)(469,928)(468,924)
\Thicklines \path(528,707)(528,707)(528,710)(523,718)(514,727)(500,738)(483,747)(465,753)(445,757)(425,758)(408,758)(394,756)(383,755)(377,754)(373,755)(369,756)(368,757)(370,760)(373,761)(378,762)(383,756)(385,743)(386,721)(387,694)(390,670)(395,652)(402,641)(410,640)(418,643)(422,651)(425,661)(425,672)(420,682)(411,689)(400,692)(388,689)(378,683)(370,674)(363,664)(359,653)(359,642)(363,629)(366,620)(369,614)(371,610)
\thinlines \path(528,707)(528,707)(529,708)(530,708)(529,709)(528,710)(526,715)(525,722)(525,730)(526,738)(528,746)(532,754)(538,759)(544,764)(551,768)(557,772)(563,776)(569,780)(576,780)(585,776)(596,765)(607,749)(609,726)(602,706)(596,694)(591,677)(587,660)(582,646)(576,633)(570,621)(566,610)(565,597)(570,588)(578,580)(587,567)(591,544)(587,512)(580,477)(575,440)(573,405)(568,379)(559,358)(552,338)(549,322)(552,310)
\thicklines \path(528,707)(528,707)(528,704)(529,702)(530,701)(530,700)(530,699)(530,697)(530,693)(529,687)(528,677)(529,667)(532,661)(535,659)(533,659)(529,664)(523,671)(515,682)(502,691)(487,700)(469,707)(449,717)(430,726)(414,734)(401,744)(393,757)(389,773)(389,788)(395,802)(405,814)(418,824)(429,831)(437,834)(439,837)(437,838)(433,841)(428,847)(423,855)(420,862)(419,867)(420,873)(422,878)(425,882)(427,884)(428,883)
\Thicklines \path(528,707)(528,707)(528,708)(529,710)(530,710)(533,709)(537,706)(543,700)(551,696)(556,691)(558,682)(561,669)(564,654)(570,636)(582,620)(598,607)(613,596)(627,587)(636,584)(646,588)(656,595)(667,602)(676,606)(683,605)(687,602)(689,598)(690,598)(688,598)(686,600)(686,602)(688,606)(691,613)(693,624)(695,640)(694,656)(691,671)(687,685)(681,700)(677,713)(670,723)(661,731)(653,737)(647,743)(644,748)(643,752)
\thinlines \path(528,707)(528,707)(527,709)(523,713)(518,719)(513,728)(506,737)(500,744)(495,749)(492,752)(494,756)(500,766)(508,781)(515,797)(524,811)(532,823)(537,831)(539,839)(539,848)(538,858)(536,872)(533,889)(532,906)(532,922)(532,936)(531,947)(529,956)(527,964)(524,972)(521,981)(518,988)(513,993)(506,996)(496,997)(486,991)(478,983)(468,973)(459,959)(460,941)(471,929)(485,929)(495,933)(500,936)(505,937)(508,933)
\thicklines \path(528,707)(528,707)(530,708)(534,713)(537,720)(539,727)(539,734)(540,739)(543,741)(548,737)(551,728)(555,716)(557,703)(560,687)(556,666)(541,647)(522,638)(503,637)(486,644)(470,657)(457,672)(450,689)(446,703)(445,716)(446,727)(447,736)(447,745)(443,755)(438,764)(436,769)(435,768)(437,764)(440,758)(441,750)(441,742)(439,733)(435,725)(429,714)(423,705)(418,694)(413,682)(411,669)(412,656)(418,644)(427,635)
\Thicklines \path(528,707)(528,707)(528,703)(526,703)(525,708)(522,709)(522,717)(520,725)(519,734)(519,742)(519,751)(517,759)(512,768)(506,777)(501,785)(501,793)(502,802)(503,810)(502,818)(502,826)(500,833)(499,841)(498,848)(497,856)(497,863)(497,870)(497,878)(498,885)(499,892)(500,899)(500,905)(501,912)(501,919)(501,925)(501,932)(500,938)(499,944)(499,951)(498,957)(498,963)(497,968)(497,974)(497,979)(496,985)(496,990)
\thinlines \path(528,707)(528,707)(528,705)(528,704)(529,704)(529,704)(531,706)(534,712)(539,721)(542,732)(544,744)(546,757)(548,773)(543,790)(536,796)(532,796)(536,792)(541,791)(547,792)(553,792)(561,793)(569,795)(576,796)(584,799)(593,799)(600,794)(606,786)(610,775)(610,762)(603,755)(594,753)(588,756)(585,762)(584,768)(583,775)(583,781)(583,788)(583,794)(583,800)(582,806)(579,811)(575,818)(575,823)(575,829)(575,834)
\thicklines \path(528,707)(528,707)(528,703)(530,695)(534,686)(539,676)(545,666)(550,661)(554,659)(556,657)(557,654)(554,653)(550,657)(548,663)(547,668)(548,672)(550,680)(550,688)(550,696)(549,704)(548,712)(546,720)(545,727)(545,734)(545,742)(545,749)(545,756)(545,764)(546,771)(547,777)(548,784)(548,790)(548,797)(548,804)(548,810)(547,817)(547,823)(546,829)(545,835)(545,841)(545,847)(544,852)(544,858)(544,864)(544,869)
\Thicklines \path(528,707)(528,707)(528,710)(523,718)(514,727)(500,738)(483,747)(465,753)(445,757)(425,758)(408,758)(394,756)(383,755)(377,754)(373,755)(369,756)(368,757)(370,760)(373,761)(378,762)(383,756)(385,743)(386,721)(387,694)(390,670)(395,652)(402,641)(410,640)(418,643)(422,651)(425,661)(425,672)(420,682)(411,689)(400,692)(388,689)(378,683)(370,674)(363,664)(359,653)(359,642)(363,629)(366,620)(369,614)(371,610)
\thinlines \path(528,707)(528,707)(526,706)(524,703)(523,698)(522,691)(521,680)(519,668)(516,659)(514,650)(514,643)(516,636)(519,629)(524,618)(529,607)(532,594)(533,581)(531,569)(527,561)(523,556)(517,554)(511,556)(503,559)(498,561)(495,565)(494,571)(495,577)(497,581)(501,581)(506,581)(512,581)(518,583)(521,587)(523,594)(525,601)(529,607)(529,613)(529,619)(528,625)(528,631)(528,637)(527,642)(528,648)(528,653)(528,659)
\end{picture}

%% file: diff
\setlength{\unitlength}{0.240900pt}
\begin{picture}(1800,900)(0,0)
\tenrm
\thicklines \path(220,113)(240,113)
\thicklines \path(1736,113)(1716,113)
\put(198,113){\makebox(0,0)[r]{0}}
\thicklines \path(220,240)(240,240)
\thicklines \path(1736,240)(1716,240)
\put(198,240){\makebox(0,0)[r]{}}
\thicklines \path(220,368)(240,368)
\thicklines \path(1736,368)(1716,368)
\put(198,368){\makebox(0,0)[r]{10}}
\thicklines \path(220,495)(240,495)
\thicklines \path(1736,495)(1716,495)
\put(198,495){\makebox(0,0)[r]{}}
\thicklines \path(220,622)(240,622)
\thicklines \path(1736,622)(1716,622)
\put(198,622){\makebox(0,0)[r]{20}}
\thicklines \path(220,750)(240,750)
\thicklines \path(1736,750)(1716,750)
\put(198,750){\makebox(0,0)[r]{}}
\thicklines \path(220,877)(240,877)
\thicklines \path(1736,877)(1716,877)
\put(198,877){\makebox(0,0)[r]{30}}
\thicklines \path(220,113)(220,133)
\thicklines \path(220,877)(220,857)
\put(220,68){\makebox(0,0){0}}
\thicklines \path(397,113)(397,133)
\thicklines \path(397,877)(397,857)
\put(397,68){\makebox(0,0){}}
\thicklines \path(575,113)(575,133)
\thicklines \path(575,877)(575,857)
\put(575,68){\makebox(0,0){40}}
\thicklines \path(752,113)(752,133)
\thicklines \path(752,877)(752,857)
\put(752,68){\makebox(0,0){}}
\thicklines \path(930,113)(930,133)
\thicklines \path(930,877)(930,857)
\put(930,68){\makebox(0,0){80}}
\thicklines \path(1107,113)(1107,133)
\thicklines \path(1107,877)(1107,857)
\put(1107,68){\makebox(0,0){}}
\thicklines \path(1285,113)(1285,133)
\thicklines \path(1285,877)(1285,857)
\put(1285,68){\makebox(0,0){120}}
\thicklines \path(1462,113)(1462,133)
\thicklines \path(1462,877)(1462,857)
\put(1462,68){\makebox(0,0){}}
\thicklines \path(1639,113)(1639,133)
\thicklines \path(1639,877)(1639,857)
\put(1639,68){\makebox(0,0){160}}
\thicklines \path(220,113)(1736,113)(1736,877)(220,877)(220,113)
\put(80,750){\makebox(0,0)[l]{\shortstack{$\langle x_*^2 \rangle$}}}
\put(978,0){\makebox(0,0){$t^*$}}
\put(237,113){\makebox(0,0){$\star$}}
\put(254,114){\makebox(0,0){$\star$}}
\put(271,115){\makebox(0,0){$\star$}}
\put(288,117){\makebox(0,0){$\star$}}
\put(305,121){\makebox(0,0){$\star$}}
\put(322,125){\makebox(0,0){$\star$}}
\put(339,131){\makebox(0,0){$\star$}}
\put(356,139){\makebox(0,0){$\star$}}
\put(373,149){\makebox(0,0){$\star$}}
\put(390,159){\makebox(0,0){$\star$}}
\put(407,171){\makebox(0,0){$\star$}}
\put(424,184){\makebox(0,0){$\star$}}
\put(441,198){\makebox(0,0){$\star$}}
\put(458,213){\makebox(0,0){$\star$}}
\put(476,227){\makebox(0,0){$\star$}}
\put(493,242){\makebox(0,0){$\star$}}
\put(510,255){\makebox(0,0){$\star$}}
\put(527,270){\makebox(0,0){$\star$}}
\put(544,284){\makebox(0,0){$\star$}}
\put(561,300){\makebox(0,0){$\star$}}
\put(578,315){\makebox(0,0){$\star$}}
\put(595,329){\makebox(0,0){$\star$}}
\put(612,342){\makebox(0,0){$\star$}}
\put(629,354){\makebox(0,0){$\star$}}
\put(646,368){\makebox(0,0){$\star$}}
\put(663,382){\makebox(0,0){$\star$}}
\put(680,397){\makebox(0,0){$\star$}}
\put(697,413){\makebox(0,0){$\star$}}
\put(714,425){\makebox(0,0){$\star$}}
\put(731,439){\makebox(0,0){$\star$}}
\put(748,454){\makebox(0,0){$\star$}}
\put(765,472){\makebox(0,0){$\star$}}
\put(782,482){\makebox(0,0){$\star$}}
\put(799,494){\makebox(0,0){$\star$}}
\put(816,509){\makebox(0,0){$\star$}}
\put(833,529){\makebox(0,0){$\star$}}
\put(850,546){\makebox(0,0){$\star$}}
\put(867,564){\makebox(0,0){$\star$}}
\put(884,578){\makebox(0,0){$\star$}}
\put(901,591){\makebox(0,0){$\star$}}
\put(918,606){\makebox(0,0){$\star$}}
\put(935,616){\makebox(0,0){$\star$}}
\put(952,628){\makebox(0,0){$\star$}}
\put(969,636){\makebox(0,0){$\star$}}
\put(987,641){\makebox(0,0){$\star$}}
\put(1004,641){\makebox(0,0){$\star$}}
\put(1021,647){\makebox(0,0){$\star$}}
\put(1038,643){\makebox(0,0){$\star$}}
\put(1055,648){\makebox(0,0){$\star$}}
\put(1072,645){\makebox(0,0){$\star$}}
\put(1089,647){\makebox(0,0){$\star$}}
\put(1106,649){\makebox(0,0){$\star$}}
\put(1123,648){\makebox(0,0){$\star$}}
\put(1140,651){\makebox(0,0){$\star$}}
\put(1157,650){\makebox(0,0){$\star$}}
\put(1174,650){\makebox(0,0){$\star$}}
\put(1191,655){\makebox(0,0){$\star$}}
\put(1208,654){\makebox(0,0){$\star$}}
\put(1225,656){\makebox(0,0){$\star$}}
\put(1242,663){\makebox(0,0){$\star$}}
\put(1259,662){\makebox(0,0){$\star$}}
\put(1276,667){\makebox(0,0){$\star$}}
\put(1293,672){\makebox(0,0){$\star$}}
\put(1310,676){\makebox(0,0){$\star$}}
\put(1327,684){\makebox(0,0){$\star$}}
\put(1344,683){\makebox(0,0){$\star$}}
\put(1361,685){\makebox(0,0){$\star$}}
\put(1378,682){\makebox(0,0){$\star$}}
\put(1395,685){\makebox(0,0){$\star$}}
\put(1412,683){\makebox(0,0){$\star$}}
\put(1429,684){\makebox(0,0){$\star$}}
\put(1446,687){\makebox(0,0){$\star$}}
\put(1463,690){\makebox(0,0){$\star$}}
\put(1480,695){\makebox(0,0){$\star$}}
\put(1498,708){\makebox(0,0){$\star$}}
\put(1515,722){\makebox(0,0){$\star$}}
\put(1532,727){\makebox(0,0){$\star$}}
\put(1549,732){\makebox(0,0){$\star$}}
\put(1566,740){\makebox(0,0){$\star$}}
\put(1583,741){\makebox(0,0){$\star$}}
\put(1600,746){\makebox(0,0){$\star$}}
\put(1617,751){\makebox(0,0){$\star$}}
\put(1634,751){\makebox(0,0){$\star$}}
\put(1651,750){\makebox(0,0){$\star$}}
\put(1668,754){\makebox(0,0){$\star$}}
\put(1685,759){\makebox(0,0){$\star$}}
\put(1702,760){\makebox(0,0){$\star$}}
\put(1719,762){\makebox(0,0){$\star$}}
\put(1736,774){\makebox(0,0){$\star$}}
\thinlines \path(340,113)(1243,877)

\put(237,113){\circle{18}}
\put(254,113){\circle{18}}
\put(271,113){\circle{18}}
\put(288,114){\circle{18}}
\put(305,115){\circle{18}}
\put(322,116){\circle{18}}
\put(339,118){\circle{18}}
\put(356,120){\circle{18}}
\put(373,123){\circle{18}}
\put(390,126){\circle{18}}
\put(407,129){\circle{18}}
\put(424,133){\circle{18}}
\put(441,137){\circle{18}}
\put(458,141){\circle{18}}
\put(476,146){\circle{18}}
\put(493,152){\circle{18}}
\put(510,158){\circle{18}}
\put(527,164){\circle{18}}
\put(544,171){\circle{18}}
\put(561,177){\circle{18}}
\put(578,183){\circle{18}}
\put(595,190){\circle{18}}
\put(612,196){\circle{18}}
\put(629,203){\circle{18}}
\put(646,210){\circle{18}}
\put(663,217){\circle{18}}
\put(680,223){\circle{18}}
\put(697,230){\circle{18}}
\put(714,238){\circle{18}}
\put(731,245){\circle{18}}
\put(748,253){\circle{18}}
\put(765,260){\circle{18}}
\put(782,269){\circle{18}}
\put(799,277){\circle{18}}
\put(816,285){\circle{18}}
\put(833,293){\circle{18}}
\put(850,299){\circle{18}}
\put(867,307){\circle{18}}
\put(884,314){\circle{18}}
\put(901,321){\circle{18}}
\put(918,329){\circle{18}}
\put(935,337){\circle{18}}
\put(952,346){\circle{18}}
\put(969,353){\circle{18}}
\put(987,359){\circle{18}}
\put(1004,365){\circle{18}}
\put(1021,371){\circle{18}}
\put(1038,375){\circle{18}}
\put(1055,379){\circle{18}}
\put(1072,381){\circle{18}}
\put(1089,383){\circle{18}}
\put(1106,385){\circle{18}}
\put(1123,387){\circle{18}}
\put(1140,387){\circle{18}}
\put(1157,387){\circle{18}}
\put(1174,388){\circle{18}}
\put(1191,389){\circle{18}}
\put(1208,389){\circle{18}}
\put(1225,391){\circle{18}}
\put(1242,391){\circle{18}}
\put(1259,389){\circle{18}}
\put(1276,392){\circle{18}}
\put(1293,392){\circle{18}}
\put(1310,394){\circle{18}}
\put(1327,398){\circle{18}}
\put(1344,403){\circle{18}}
\put(1361,406){\circle{18}}
\put(1378,410){\circle{18}}
\put(1395,416){\circle{18}}
\put(1412,423){\circle{18}}
\put(1429,425){\circle{18}}
\put(1446,431){\circle{18}}
\put(1463,433){\circle{18}}
\put(1480,437){\circle{18}}
\put(1498,442){\circle{18}}
\put(1515,448){\circle{18}}
\put(1532,450){\circle{18}}
\put(1549,451){\circle{18}}
\put(1566,456){\circle{18}}
\put(1583,460){\circle{18}}
\put(1600,465){\circle{18}}
\put(1617,470){\circle{18}}
\put(1634,473){\circle{18}}
\put(1651,478){\circle{18}}
\put(1668,480){\circle{18}}
\put(1685,484){\circle{18}}
\put(1702,489){\circle{18}}
\put(1719,490){\circle{18}}
\put(1736,495){\circle{18}}
\thinlines \path(415,113)(1736,682)

\end{picture}

%% file: sed_new1b
\setlength{\unitlength}{0.240900pt}
\begin{picture}(1049,900)(0,0)
\tenrm
\thicklines \path(220,113)(240,113)
\thicklines \path(985,113)(965,113)
\put(198,113){\makebox(0,0)[r]{0}}
\thicklines \path(220,304)(240,304)
\thicklines \path(985,304)(965,304)
\put(198,304){\makebox(0,0)[r]{}}
\thicklines \path(220,495)(240,495)
\thicklines \path(985,495)(965,495)
\put(198,495){\makebox(0,0)[r]{10}}
\thicklines \path(220,686)(240,686)
\thicklines \path(985,686)(965,686)
\put(198,686){\makebox(0,0)[r]{}}
\thicklines \path(220,877)(240,877)
\thicklines \path(985,877)(965,877)
\put(198,877){\makebox(0,0)[r]{20}}
\thicklines \path(220,113)(220,133)
\thicklines \path(220,877)(220,857)
\put(220,68){\makebox(0,0){0.0}}
\thicklines \path(310,113)(310,133)
\thicklines \path(310,877)(310,857)
\put(310,68){\makebox(0,0){}}
\thicklines \path(400,113)(400,133)
\thicklines \path(400,877)(400,857)
\put(400,68){\makebox(0,0){0.1}}
\thicklines \path(490,113)(490,133)
\thicklines \path(490,877)(490,857)
\put(490,68){\makebox(0,0){}}
\thicklines \path(580,113)(580,133)
\thicklines \path(580,877)(580,857)
\put(580,68){\makebox(0,0){0.2}}
\thicklines \path(670,113)(670,133)
\thicklines \path(670,877)(670,857)
\put(670,68){\makebox(0,0){}}
\thicklines \path(760,113)(760,133)
\thicklines \path(760,877)(760,857)
\put(760,68){\makebox(0,0){0.3}}
\thicklines \path(850,113)(850,133)
\thicklines \path(850,877)(850,857)
\put(850,68){\makebox(0,0){}}
\thicklines \path(940,113)(940,133)
\thicklines \path(940,877)(940,857)
\put(940,68){\makebox(0,0){0.4}}
\thicklines \path(220,113)(985,113)(985,877)(220,877)(220,113)
\put(80,686){\makebox(0,0)[l]{\shortstack{$D^*$}}}
\put(602,0){\makebox(0,0){$\Phi$}}
\put(895,801){\makebox(0,0){\bf (a)}}
\put(255,461){\circle{18}}
\put(291,532){\circle{18}}
\put(326,649){\circle{18}}
\put(361,737){\circle{18}}
\put(431,761){\circle{18}}
\put(503,688){\circle{18}}
\put(573,720){\circle{18}}
\put(643,640){\circle{18}}
\put(713,573){\circle{18}}
\put(785,510){\circle{18}}
\put(855,364){\circle{18}}
\put(927,311){\circle{18}}
\thinlines \path(255,442)(255,481)
\thinlines \path(245,442)(265,442)
\thinlines \path(245,481)(265,481)
\thinlines \path(291,532)(291,532)
\thinlines \path(281,532)(301,532)
\thinlines \path(281,532)(301,532)
\thinlines \path(326,595)(326,704)
\thinlines \path(316,595)(336,595)
\thinlines \path(316,704)(336,704)
\thinlines \path(361,737)(361,737)
\thinlines \path(351,737)(371,737)
\thinlines \path(351,737)(371,737)
\thinlines \path(431,687)(431,835)
\thinlines \path(421,687)(441,687)
\thinlines \path(421,835)(441,835)
\thinlines \path(503,570)(503,806)
\thinlines \path(493,570)(513,570)
\thinlines \path(493,806)(513,806)
\thinlines \path(573,720)(573,720)
\thinlines \path(563,720)(583,720)
\thinlines \path(563,720)(583,720)
\thinlines \path(643,640)(643,640)
\thinlines \path(633,640)(653,640)
\thinlines \path(633,640)(653,640)
\thinlines \path(713,528)(713,618)
\thinlines \path(703,528)(723,528)
\thinlines \path(703,618)(723,618)
\thinlines \path(785,408)(785,613)
\thinlines \path(775,408)(795,408)
\thinlines \path(775,613)(795,613)
\thinlines \path(855,364)(855,364)
\thinlines \path(845,364)(865,364)
\thinlines \path(845,364)(865,364)
\thinlines \path(927,258)(927,363)
\thinlines \path(917,258)(937,258)
\thinlines \path(917,363)(937,363)
\thinlines \path(255,172)(255,172)(291,203)(326,229)(361,261)(431,280)(503,279)(573,300)(643,266)
\thinlines \path(643,266)(713,266)(785,245)(855,227)(927,192)
\put(255,172){\makebox(0,0){$\star$}}
\put(291,203){\makebox(0,0){$\star$}}
\put(326,229){\makebox(0,0){$\star$}}
\put(361,261){\makebox(0,0){$\star$}}
\put(431,280){\makebox(0,0){$\star$}}
\put(503,279){\makebox(0,0){$\star$}}
\put(573,300){\makebox(0,0){$\star$}}
\put(643,266){\makebox(0,0){$\star$}}
\put(713,266){\makebox(0,0){$\star$}}
\put(785,245){\makebox(0,0){$\star$}}
\put(855,227){\makebox(0,0){$\star$}}
\put(927,192){\makebox(0,0){$\star$}}
\thinlines \path(255,461)(255,461)(291,532)(326,649)(361,737)(431,761)(503,688)(573,720)(643,640)
\thinlines \path(643,640)(713,573)(785,510)(855,364)(927,311)
\end{picture}

%% file: sed_new1c
\setlength{\unitlength}{0.240900pt}
\begin{picture}(1049,900)(0,0)
\tenrm
\thicklines \path(220,113)(240,113)
\thicklines \path(985,113)(965,113)
\put(198,113){\makebox(0,0)[r]{0}}
\thicklines \path(220,222)(240,222)
\thicklines \path(985,222)(965,222)
\put(198,222){\makebox(0,0)[r]{}}
\thicklines \path(220,331)(240,331)
\thicklines \path(985,331)(965,331)
\put(198,331){\makebox(0,0)[r]{2}}
\thicklines \path(220,440)(240,440)
\thicklines \path(985,440)(965,440)
\put(198,440){\makebox(0,0)[r]{}}
\thicklines \path(220,550)(240,550)
\thicklines \path(985,550)(965,550)
\put(198,550){\makebox(0,0)[r]{4}}
\thicklines \path(220,659)(240,659)
\thicklines \path(985,659)(965,659)
\put(198,659){\makebox(0,0)[r]{}}
\thicklines \path(220,768)(240,768)
\thicklines \path(985,768)(965,768)
\put(198,768){\makebox(0,0)[r]{6}}
\thicklines \path(220,877)(240,877)
\thicklines \path(985,877)(965,877)
\put(198,877){\makebox(0,0)[r]{}}
\thicklines \path(220,113)(220,133)
\thicklines \path(220,877)(220,857)
\put(220,68){\makebox(0,0){0.0}}
\thicklines \path(310,113)(310,133)
\thicklines \path(310,877)(310,857)
\put(310,68){\makebox(0,0){}}
\thicklines \path(400,113)(400,133)
\thicklines \path(400,877)(400,857)
\put(400,68){\makebox(0,0){0.1}}
\thicklines \path(490,113)(490,133)
\thicklines \path(490,877)(490,857)
\put(490,68){\makebox(0,0){}}
\thicklines \path(580,113)(580,133)
\thicklines \path(580,877)(580,857)
\put(580,68){\makebox(0,0){0.2}}
\thicklines \path(670,113)(670,133)
\thicklines \path(670,877)(670,857)
\put(670,68){\makebox(0,0){}}
\thicklines \path(760,113)(760,133)
\thicklines \path(760,877)(760,857)
\put(760,68){\makebox(0,0){0.3}}
\thicklines \path(850,113)(850,133)
\thicklines \path(850,877)(850,857)
\put(850,68){\makebox(0,0){}}
\thicklines \path(940,113)(940,133)
\thicklines \path(940,877)(940,857)
\put(940,68){\makebox(0,0){0.4}}
\thicklines \path(220,113)(985,113)(985,877)(220,877)(220,113)
\put(80,659){\makebox(0,0)[l]{\shortstack{\Large $\frac{D_y^*}{D_x^*}$}}}
\put(602,0){\makebox(0,0){$\Phi$}}
\put(895,801){\makebox(0,0){\bf (b)}}
\thinlines \path(255,763)(255,763)(291,619)(326,619)(361,573)(431,536)(503,491)(573,467)(643,489)
\thinlines \path(643,489)(713,440)(785,441)(855,353)(927,385)
\put(255,763){\circle{18}}
\put(291,619){\circle{18}}
\put(326,619){\circle{18}}
\put(361,573){\circle{18}}
\put(431,536){\circle{18}}
\put(503,491){\circle{18}}
\put(573,467){\circle{18}}
\put(643,489){\circle{18}}
\put(713,440){\circle{18}}
\put(785,441){\circle{18}}
\put(855,353){\circle{18}}
\put(927,385){\circle{18}}
\put(310,671){\makebox(0,0){$+$}}
\put(400,812){\makebox(0,0){$+$}}
\put(490,623){\makebox(0,0){$+$}}
\put(580,681){\makebox(0,0){$+$}}
\put(670,518){\makebox(0,0){$+$}}
\put(760,550){\makebox(0,0){$+$}}
\put(850,529){\makebox(0,0){$+$}}
\put(940,522){\makebox(0,0){$+$}}
\end{picture}

%% file: sed_new3a
\setlength{\unitlength}{0.240900pt}
\begin{picture}(1049,900)(0,0)
\tenrm
\thicklines \path(220,113)(240,113)
\thicklines \path(985,113)(965,113)
\put(198,113){\makebox(0,0)[r]{0.0}}
\thicklines \path(220,204)(240,204)
\thicklines \path(985,204)(965,204)
\put(198,204){\makebox(0,0)[r]{}}
\thicklines \path(220,295)(240,295)
\thicklines \path(985,295)(965,295)
\put(198,295){\makebox(0,0)[r]{0.25}}
\thicklines \path(220,386)(240,386)
\thicklines \path(985,386)(965,386)
\put(198,386){\makebox(0,0)[r]{}}
\thicklines \path(220,477)(240,477)
\thicklines \path(985,477)(965,477)
\put(198,477){\makebox(0,0)[r]{0.5}}
\thicklines \path(220,568)(240,568)
\thicklines \path(985,568)(965,568)
\put(198,568){\makebox(0,0)[r]{}}
\thicklines \path(220,659)(240,659)
\thicklines \path(985,659)(965,659)
\put(198,659){\makebox(0,0)[r]{0.75}}
\thicklines \path(220,750)(240,750)
\thicklines \path(985,750)(965,750)
\put(198,750){\makebox(0,0)[r]{}}
\thicklines \path(220,841)(240,841)
\thicklines \path(985,841)(965,841)
\put(198,841){\makebox(0,0)[r]{1.0}}
\thicklines \path(220,113)(220,133)
\thicklines \path(220,877)(220,857)
\put(220,68){\makebox(0,0){0.0}}
\thicklines \path(310,113)(310,133)
\thicklines \path(310,877)(310,857)
\put(310,68){\makebox(0,0){}}
\thicklines \path(400,113)(400,133)
\thicklines \path(400,877)(400,857)
\put(400,68){\makebox(0,0){0.1}}
\thicklines \path(490,113)(490,133)
\thicklines \path(490,877)(490,857)
\put(490,68){\makebox(0,0){}}
\thicklines \path(580,113)(580,133)
\thicklines \path(580,877)(580,857)
\put(580,68){\makebox(0,0){0.2}}
\thicklines \path(670,113)(670,133)
\thicklines \path(670,877)(670,857)
\put(670,68){\makebox(0,0){}}
\thicklines \path(760,113)(760,133)
\thicklines \path(760,877)(760,857)
\put(760,68){\makebox(0,0){0.3}}
\thicklines \path(850,113)(850,133)
\thicklines \path(850,877)(850,857)
\put(850,68){\makebox(0,0){}}
\thicklines \path(940,113)(940,133)
\thicklines \path(940,877)(940,857)
\put(940,68){\makebox(0,0){0.4}}
\thicklines \path(220,113)(985,113)(985,877)(220,877)(220,113)
\put(80,750){\makebox(0,0)[l]{\shortstack{\Large $\frac{\langle V \rangle}{V_S}$}}}
\put(602,0){\makebox(0,0){$\Phi$}}
\put(895,801){\makebox(0,0){\bf (a)}}
\put(255,783){\circle{18}}
\put(291,724){\circle{18}}
\put(326,677){\circle{18}}
\put(361,650){\circle{18}}
\put(431,605){\circle{18}}
\put(503,564){\circle{18}}
\put(573,541){\circle{18}}
\put(643,504){\circle{18}}
\put(713,478){\circle{18}}
\put(785,459){\circle{18}}
\put(855,429){\circle{18}}
\put(927,408){\circle{18}}
\thinlines \path(220,841)(220,841)(228,831)(235,822)(243,813)(251,804)(259,795)(266,786)(274,777)
\thinlines \path(274,777)(282,768)(290,759)(297,751)(305,742)(313,734)(320,725)(328,717)(336,709)
\thinlines \path(336,709)(344,701)(351,693)(359,685)(367,677)(375,669)(382,661)(390,653)(398,646)
\thinlines \path(398,646)(405,638)(413,631)(421,623)(429,616)(436,609)(444,601)(452,594)(460,587)
\thinlines \path(460,587)(467,580)(475,573)(483,566)(490,559)(498,553)(506,546)(514,539)(521,533)
\thinlines \path(521,533)(529,526)(537,520)(545,514)(552,507)(560,501)(568,495)(575,489)(583,483)
\thinlines \path(583,483)(591,477)(599,471)(606,465)(614,460)(622,454)(630,448)(637,443)(645,437)
\thinlines \path(645,437)(653,432)(660,427)(668,421)(676,416)(684,411)(691,406)(699,401)(707,396)
\thinlines \path(707,396)(715,391)(722,386)(730,381)(738,376)(745,371)(753,367)(761,362)(769,357)
\thinlines \path(769,357)(776,353)(784,349)(792,344)(800,340)(807,336)(815,331)(823,327)(830,323)
\thinlines \path(830,323)(838,319)(846,315)(854,311)(861,307)(869,303)(877,299)(885,296)(892,292)
\thinlines \path(892,292)(900,288)(908,285)(915,281)(923,278)(931,274)(939,271)(946,267)(954,264)
\thinlines \path(954,264)(962,261)(970,258)(977,254)(985,251)
\put(255,820){\makebox(0,0){$\star$}}
\put(291,730){\makebox(0,0){$\star$}}
\put(326,724){\makebox(0,0){$\star$}}
\put(361,686){\makebox(0,0){$\star$}}
\put(432,637){\makebox(0,0){$\star$}}
\put(503,564){\makebox(0,0){$\star$}}
\put(573,531){\makebox(0,0){$\star$}}
\put(644,502){\makebox(0,0){$\star$}}
\put(715,446){\makebox(0,0){$\star$}}
\put(786,415){\makebox(0,0){$\star$}}
\put(856,387){\makebox(0,0){$\star$}}
\put(927,364){\makebox(0,0){$\star$}}
\end{picture}

%% file: sed_new3d
\setlength{\unitlength}{0.240900pt}
\begin{picture}(1049,900)(0,0)
\tenrm
\thicklines \path(220,113)(240,113)
\thicklines \path(985,113)(965,113)
\put(198,113){\makebox(0,0)[r]{0.0}}
\thicklines \path(220,304)(240,304)
\thicklines \path(985,304)(965,304)
\put(198,304){\makebox(0,0)[r]{}}
\thicklines \path(220,495)(240,495)
\thicklines \path(985,495)(965,495)
\put(198,495){\makebox(0,0)[r]{0.5}}
\thicklines \path(220,686)(240,686)
\thicklines \path(985,686)(965,686)
\put(198,686){\makebox(0,0)[r]{}}
\thicklines \path(220,877)(240,877)
\thicklines \path(985,877)(965,877)
\put(198,877){\makebox(0,0)[r]{1.0}}
\thicklines \path(220,113)(220,133)
\thicklines \path(220,877)(220,857)
\put(220,68){\makebox(0,0){0}}
\thicklines \path(329,113)(329,133)
\thicklines \path(329,877)(329,857)
\put(329,68){\makebox(0,0){}}
\thicklines \path(439,113)(439,133)
\thicklines \path(439,877)(439,857)
\put(439,68){\makebox(0,0){1}}
\thicklines \path(548,113)(548,133)
\thicklines \path(548,877)(548,857)
\put(548,68){\makebox(0,0){}}
\thicklines \path(657,113)(657,133)
\thicklines \path(657,877)(657,857)
\put(657,68){\makebox(0,0){2}}
\thicklines \path(766,113)(766,133)
\thicklines \path(766,877)(766,857)
\put(766,68){\makebox(0,0){}}
\thicklines \path(876,113)(876,133)
\thicklines \path(876,877)(876,857)
\put(876,68){\makebox(0,0){3}}
\thicklines \path(985,113)(985,133)
\thicklines \path(985,877)(985,857)
\put(985,68){\makebox(0,0){}}
\thicklines \path(220,113)(985,113)(985,877)(220,877)(220,113)
\put(80,686){\makebox(0,0)[l]{\shortstack{\Large $\frac{\langle V \rangle}{V_S}$}}}
\put(602,0){\makebox(0,0){\Large $\frac{\Delta x}{\bar{r}}$}}
\put(895,801){\makebox(0,0){\bf (b)}}
\thinlines \path(948,721)(948,721)(766,677)(657,693)(585,689)(493,715)
\thinlines \path(948,614)(948,614)(766,587)(657,575)(585,574)(493,587)
\thinlines \path(948,578)(948,578)(766,523)(657,508)(585,509)(493,521)
\thinlines \path(948,524)(948,524)(766,477)(657,445)(585,430)(493,431)
\thinlines \path(948,481)(948,481)(766,422)(657,392)(585,378)(493,376)
\put(948,721){\circle{18}}
\put(766,677){\circle{18}}
\put(657,693){\circle{18}}
\put(585,689){\circle{18}}
\put(493,715){\circle{18}}
\put(948,614){\circle{18}}
\put(766,587){\circle{18}}
\put(657,575){\circle{18}}
\put(585,574){\circle{18}}
\put(493,587){\circle{18}}
\put(948,578){\circle{18}}
\put(766,523){\circle{18}}
\put(657,508){\circle{18}}
\put(585,509){\circle{18}}
\put(493,521){\circle{18}}
\put(948,524){\circle{18}}
\put(766,477){\circle{18}}
\put(657,445){\circle{18}}
\put(585,430){\circle{18}}
\put(493,431){\circle{18}}
\put(948,481){\circle{18}}
\put(766,422){\circle{18}}
\put(657,392){\circle{18}}
\put(585,378){\circle{18}}
\put(493,376){\circle{18}}
\end{picture}

%% file: sed_new4a
\setlength{\unitlength}{0.240900pt}
\begin{picture}(1049,900)(0,0)
\tenrm
\thicklines \path(220,113)(240,113)
\thicklines \path(985,113)(965,113)
\put(198,113){\makebox(0,0)[r]{0.0}}
\thicklines \path(220,304)(240,304)
\thicklines \path(985,304)(965,304)
\put(198,304){\makebox(0,0)[r]{}}
\thicklines \path(220,495)(240,495)
\thicklines \path(985,495)(965,495)
\put(198,495){\makebox(0,0)[r]{0.5}}
\thicklines \path(220,686)(240,686)
\thicklines \path(985,686)(965,686)
\put(198,686){\makebox(0,0)[r]{}}
\thicklines \path(220,877)(240,877)
\thicklines \path(985,877)(965,877)
\put(198,877){\makebox(0,0)[r]{1.0}}
\thicklines \path(220,113)(220,133)
\thicklines \path(220,877)(220,857)
\put(220,68){\makebox(0,0){0.0}}
\thicklines \path(310,113)(310,133)
\thicklines \path(310,877)(310,857)
\put(310,68){\makebox(0,0){}}
\thicklines \path(400,113)(400,133)
\thicklines \path(400,877)(400,857)
\put(400,68){\makebox(0,0){0.1}}
\thicklines \path(490,113)(490,133)
\thicklines \path(490,877)(490,857)
\put(490,68){\makebox(0,0){}}
\thicklines \path(580,113)(580,133)
\thicklines \path(580,877)(580,857)
\put(580,68){\makebox(0,0){0.2}}
\thicklines \path(670,113)(670,133)
\thicklines \path(670,877)(670,857)
\put(670,68){\makebox(0,0){}}
\thicklines \path(760,113)(760,133)
\thicklines \path(760,877)(760,857)
\put(760,68){\makebox(0,0){0.3}}
\thicklines \path(850,113)(850,133)
\thicklines \path(850,877)(850,857)
\put(850,68){\makebox(0,0){}}
\thicklines \path(940,113)(940,133)
\thicklines \path(940,877)(940,857)
\put(940,68){\makebox(0,0){0.4}}
\thicklines \path(220,113)(985,113)(985,877)(220,877)(220,113)
\put(80,686){\makebox(0,0)[l]{\shortstack{$\Delta v_y^*$}}}
\put(602,0){\makebox(0,0){$\Phi$}}
\put(895,801){\makebox(0,0){\bf (a)}}
\put(255,435){\circle{18}}
\put(291,533){\circle{18}}
\put(326,538){\circle{18}}
\put(361,635){\circle{18}}
\put(431,668){\circle{18}}
\put(503,632){\circle{18}}
\put(573,706){\circle{18}}
\put(643,675){\circle{18}}
\put(713,624){\circle{18}}
\put(785,623){\circle{18}}
\put(855,542){\circle{18}}
\put(927,448){\circle{18}}
\thinlines \path(255,384)(255,487)
\thinlines \path(245,384)(265,384)
\thinlines \path(245,487)(265,487)
\thinlines \path(291,533)(291,533)
\thinlines \path(281,533)(301,533)
\thinlines \path(281,533)(301,533)
\thinlines \path(326,519)(326,557)
\thinlines \path(316,519)(336,519)
\thinlines \path(316,557)(336,557)
\thinlines \path(361,635)(361,635)
\thinlines \path(351,635)(371,635)
\thinlines \path(351,635)(371,635)
\thinlines \path(431,639)(431,696)
\thinlines \path(421,639)(441,639)
\thinlines \path(421,696)(441,696)
\thinlines \path(503,584)(503,680)
\thinlines \path(493,584)(513,584)
\thinlines \path(493,680)(513,680)
\thinlines \path(573,706)(573,706)
\thinlines \path(563,706)(583,706)
\thinlines \path(563,706)(583,706)
\thinlines \path(643,675)(643,675)
\thinlines \path(633,675)(653,675)
\thinlines \path(633,675)(653,675)
\thinlines \path(713,608)(713,640)
\thinlines \path(703,608)(723,608)
\thinlines \path(703,640)(723,640)
\thinlines \path(785,589)(785,658)
\thinlines \path(775,589)(795,589)
\thinlines \path(775,658)(795,658)
\thinlines \path(855,542)(855,542)
\thinlines \path(845,542)(865,542)
\thinlines \path(845,542)(865,542)
\thinlines \path(927,422)(927,474)
\thinlines \path(917,422)(937,422)
\thinlines \path(917,474)(937,474)
\thinlines \path(255,435)(255,435)(291,533)(326,538)(361,635)(431,668)(503,632)(573,706)(643,675)
\thinlines \path(643,675)(713,624)(785,623)(855,542)(927,448)
\put(524,633){\makebox(0,0){$\times$}}
\put(603,489){\makebox(0,0){$\times$}}
\put(686,393){\makebox(0,0){$\times$}}
\put(749,290){\makebox(0,0){$\times$}}
\put(852,220){\makebox(0,0){$\times$}}
\put(927,145){\makebox(0,0){$\times$}}
\thinlines \path(524,600)(524,665)
\thinlines \path(514,600)(534,600)
\thinlines \path(514,665)(534,665)
\thinlines \path(603,471)(603,507)
\thinlines \path(593,471)(613,471)
\thinlines \path(593,507)(613,507)
\thinlines \path(686,377)(686,409)
\thinlines \path(676,377)(696,377)
\thinlines \path(676,409)(696,409)
\thinlines \path(749,274)(749,307)
\thinlines \path(739,274)(759,274)
\thinlines \path(739,307)(759,307)
\thinlines \path(852,187)(852,252)
\thinlines \path(842,187)(862,187)
\thinlines \path(842,252)(862,252)
\thinlines \path(927,113)(927,177)
\thinlines \path(917,113)(937,113)
\thinlines \path(917,177)(937,177)
\put(310,587){\makebox(0,0){$+$}}
\put(400,633){\makebox(0,0){$+$}}
\put(490,579){\makebox(0,0){$+$}}
\put(580,449){\makebox(0,0){$+$}}
\put(670,411){\makebox(0,0){$+$}}
\put(760,335){\makebox(0,0){$+$}}
\put(850,258){\makebox(0,0){$+$}}
\put(940,182){\makebox(0,0){$+$}}
\end{picture}

%% file: sed_new4b
\setlength{\unitlength}{0.240900pt}
\begin{picture}(1049,900)(0,0)
\tenrm
\thicklines \path(220,113)(240,113)
\thicklines \path(985,113)(965,113)
\put(198,113){\makebox(0,0)[r]{}}
\thicklines \path(220,240)(240,240)
\thicklines \path(985,240)(965,240)
\put(198,240){\makebox(0,0)[r]{1.0}}
\thicklines \path(220,368)(240,368)
\thicklines \path(985,368)(965,368)
\put(198,368){\makebox(0,0)[r]{}}
\thicklines \path(220,495)(240,495)
\thicklines \path(985,495)(965,495)
\put(198,495){\makebox(0,0)[r]{1.5}}
\thicklines \path(220,622)(240,622)
\thicklines \path(985,622)(965,622)
\put(198,622){\makebox(0,0)[r]{}}
\thicklines \path(220,750)(240,750)
\thicklines \path(985,750)(965,750)
\put(198,750){\makebox(0,0)[r]{2.0}}
\thicklines \path(220,877)(240,877)
\thicklines \path(985,877)(965,877)
\put(198,877){\makebox(0,0)[r]{}}
\thicklines \path(220,113)(220,133)
\thicklines \path(220,877)(220,857)
\put(220,68){\makebox(0,0){0.0}}
\thicklines \path(310,113)(310,133)
\thicklines \path(310,877)(310,857)
\put(310,68){\makebox(0,0){}}
\thicklines \path(400,113)(400,133)
\thicklines \path(400,877)(400,857)
\put(400,68){\makebox(0,0){0.1}}
\thicklines \path(490,113)(490,133)
\thicklines \path(490,877)(490,857)
\put(490,68){\makebox(0,0){}}
\thicklines \path(580,113)(580,133)
\thicklines \path(580,877)(580,857)
\put(580,68){\makebox(0,0){0.2}}
\thicklines \path(670,113)(670,133)
\thicklines \path(670,877)(670,857)
\put(670,68){\makebox(0,0){}}
\thicklines \path(760,113)(760,133)
\thicklines \path(760,877)(760,857)
\put(760,68){\makebox(0,0){0.3}}
\thicklines \path(850,113)(850,133)
\thicklines \path(850,877)(850,857)
\put(850,68){\makebox(0,0){}}
\thicklines \path(940,113)(940,133)
\thicklines \path(940,877)(940,857)
\put(940,68){\makebox(0,0){0.4}}
\thicklines \path(220,113)(985,113)(985,877)(220,877)(220,113)
\put(80,622){\makebox(0,0)[l]{\shortstack{\Large $\frac{\Delta v_y^*}{\Delta v_x^*}$}}}
\put(602,0){\makebox(0,0){$\Phi$}}
\put(895,801){\makebox(0,0){\bf (b)}}
\thinlines \path(255,415)(255,415)(291,382)(326,444)(361,429)(431,513)(503,435)(573,469)(643,408)
\thinlines \path(643,408)(713,404)(785,412)(855,425)(927,388)
\put(255,415){\circle{18}}
\put(291,382){\circle{18}}
\put(326,444){\circle{18}}
\put(361,429){\circle{18}}
\put(431,513){\circle{18}}
\put(503,435){\circle{18}}
\put(573,469){\circle{18}}
\put(643,408){\circle{18}}
\put(713,404){\circle{18}}
\put(785,412){\circle{18}}
\put(855,425){\circle{18}}
\put(927,388){\circle{18}}
\put(310,689){\makebox(0,0){$+$}}
\put(400,719){\makebox(0,0){$+$}}
\put(490,699){\makebox(0,0){$+$}}
\put(580,663){\makebox(0,0){$+$}}
\put(670,561){\makebox(0,0){$+$}}
\put(760,510){\makebox(0,0){$+$}}
\put(850,424){\makebox(0,0){$+$}}
\put(940,388){\makebox(0,0){$+$}}
\end{picture}